# Graphite superlubricity enabled by triboinduced nanocontacts

Renato Buzio[1]*, Andrea Gerbi[1], Cristina Bernini[1], Luca Repetto[2], Andrea Vanossi[3,4]

[1]CNR-SPIN, C.so F.M. Perrone 24, 16152 Genova, Italy

[2]Dipartimento di Fisica, Università degli Studi di Genova, Via Dodecaneso 33, 16146 Genova, Italy

[3]International School for Advanced Studies (SISSA), Via Bonomea 265, 34136 Trieste, Italy

[4]CNR-IOM Democritos National Simulation Center, Via Bonomea 265, 34136 Trieste, Italy

Colloidal probe Atomic Force Microscopy allows to explore sliding states of vanishing friction, i.e. superlubricity, in mesoscopic graphite contacts. Superlubricity is known to appear upon formation of a triboinduced transfer layer, originated by material transfer of graphene flakes from the graphitic substrate to the colloidal probe. Previous studies suggest that friction vanishes due to crystalline incommensurability at the newly formed interface. However this picture still lacks several details, such as the roles of the tribolayer roughness and of loading conditions. Hereafter we gain deeper insight into the tribological response of micrometric silica beads sliding on graphite under ambient conditions. We show that the tribotransferred flakes behave as lubricious nanoasperities with a twofold role. First, they decrease the silica-graphite true contact area, in fact causing a breakdown of adhesion and friction by one order of magnitude. Second, they govern mechanical dissipation through the specific energy landscape experienced by the topographically-highest triboinduced nanoasperity. Remarkably, such contact junctions can undergo a load-driven atomic-scale transition from continuous superlubric sliding to dissipative stick-slip, that agrees with the single-asperity Prandtl-Tomlinson model. Superlubricity in mesoscopic silica-graphite junctions may therefore arise from the load-controlled competition between interfacial crystalline incommensurability and contact pinning effects at one dominant nanoasperity.

*Corresponding author. Tel.: +39 010 6598 731. E-mail: renato.buzio@spin.cnr.it (Renato Buzio)

**Keywords:** colloidal probe AFM, graphene, transfer layer, roughness, superlubricity, atomic friction.

## 1. INTRODUCTION

Superlubricity is an umbrella term for situations in which sliding friction vanishes or very nearly vanishes [1]. Different strategies can be implemented to its achievement, involving both dry and liquid lubrication. Graphitic systems such as bulk graphite and graphene flakes [2–6], graphene nanoribbons [7–9], fullerenes [10] and nanotubes [11–13], whose interactions are dominated by the weak van der Waals forces, are well known to support dry superlubricity thanks to the formation of easy share planes across atomically smooth contact interfaces [14]. Specifically for bulk graphite, experiments from macro- to nano-scale have demonstrated that dry superlubricity appears after the built-in of a transfer layer (TL), derived by triboinduced material transfer from the graphitic substrate to the sliding countersurface [5,15–19]. In analogy with other solid lamellar lubricants (e.g. $MoS_2$ [20]) the graphitic TL mostly consists of nanosized flakes [19,21], that are thought to arrange in misfit configurations thus providing interfacial crystalline incommensurability and friction coefficients as small as $10^{-3}$. This superlow friction state, called structural lubricity, represents a peculiar condition characterized by suppression of the energy barriers to sliding motion due to cancellation of the atomistic lateral forces in an incommensurate crystalline interface. Direct evidences for structural lubricity in single-asperity graphite contacts were obtained at the nanoscale by Dienwiebel *et al.* [5], using a sharp Atomic Force Microscopy (AFM) probe sliding on graphite. In this case the TL consisted of an individual nanoflake transferred from graphite to the AFM probe tip [19]. Structural lubricity was later recognized in other nanoscale graphitic systems, namely carbon nanotubes, nanoribbons and fullerene-based nanobearings [22]. These successful results have pushed tribological research into the attempt to engineer structural lubricity from the nano- to micro- and even to macro-scale, by lubrication via selected graphite-based nanomaterials. Colloidal probe AFM experiments involving smooth micrometric beads represent a powerful tool to study the scale up of structural lubricity in mesoscopic graphite tribopairs. In fact in AFM-based methods, the graphitic TL retains to a good approximation the crystalline phase, layered structure and elemental composition of inert graphite [5,19,23–25]. Severe mechanical or chemical degradation of the tribotransferred flakes (due tribochemical reactions or environment effects) are not dominating factors, as found for other testing conditions and superlubric 2D materials [26–30]. In particular it has been shown that when a silica bead slides on Highly Oriented Pyrolitic Graphite (HOPG) under ambient conditions, the contact may evolve towards a low-adhesion, low-friction state due to the formation of a TL on the silica bead. The TL consists of few-layer/multilayer graphene flakes [24]. Friction coefficients as small as $3 \times 10^{-4}$ have been measured in a limited range of normal loads ($\leq 400\text{nN}$). It was speculated that structural lubricity relies in such case on the roughness of the sphere surface, that incorporates elevated nanoasperities covered by randomly oriented graphene nanopatches. Such a multi-asperity interfacial geometry involves the cumulative effect of many incommensurate nanocontacts and is expected to be robust against variations of load, humidity or orientational effects. This scenario holds in principle great promises, as suggests a conceptual route to extend structural lubricity to macroscopic rough contacts [3]. At present, however, detailed knowledge is missing of the actual mechanisms that assist (or suppress) the organization of the tribotransferred flakes into a robust, incommensurate rough contact.

In this work we deeply characterize the tribological response of the graphitic TL originated in colloidal AFM friction experiments. We originally complement friction force spectroscopy with knowledge derived from TL morphology, AFM convolution effects and atomically resolved friction maps. In this way we achieve a detailed description of the contact mechanics and we identify, thanks also to numerical modeling, specific load-dependent mechanisms governing different friction regimes (including thermo- and super-lubricity) for such tribosystem. All evidences ascribe a key role to the TL nanoroughness: it makes the colloidal bead-HOPG contact nanoscopic in size, and thus it enables the appearance of the structural superlubricity mechanism typically encountered in nanoscale graphitic systems. This picture is prone to generalization to mesoscale contacts with other 2D materials.

## 2. Materials and methods

**2.1 Graphite samples.** We used a square HOPG substrate of grade ZYB (MikroMasch; $10 \times 10\text{mm}^2$ in size, $0.8° \pm 0.2°$ mosaic spread). The average grains size, as directly visualized by Scanning Electron Microscopy (SEM), was $\approx 6.6\mu\text{m}$ [31]. According to the manufacturer (private communication), the armchair and zigzag crystallographic directions of the (0001) HOPG surface are parallel to the edges of the square substrate. Graphite was freshly cleaved by adhesive tape in ambient air prior to AFM measurements. The substrate was usually mounted onto the AFM sample holder with a fixed orientation, that resulted in an AFM sliding direction (approximately) perpendicular to the surface-exposed atomic steps (see Supplementary Data section S1). This choice aids material transfer from HOPG to the pristine colloidal probes (compared to other substrate orientations). We realized *a posteriori*, by analyzing atomic-scale friction maps, that such a substrate orientation corresponded to a sliding direction slightly off ($\leq 10°$) the armchair crystallographic direction (see section 3.3 for details). A few friction experiments were also carried out on a different HOPG substrate of higher crystalline quality, i.e. grade ZYA (MikroMasch, $0.4° \pm 0.1°$ mosaic spread). However, we did not observe any correlation of the friction phenomenology with the crystalline quality of the substrates.

**2.2 Morphological and friction measurements.** AFM experiments were performed by means of a commercial AFM (Solver P47-PRO by NT-MDT, Russia) operated in contact mode under ambient conditions (relative humidity RH = 30% - 60%, temperature T=23±3°C). Relative Humidity (RH) and temperature were monitored by means of a thermo-hygrometer (605-H1 by Testo). RH was stable within $\pm 1\%$ for several hours, which is largely above the timescale for individual friction force spectroscopy experiments. Large fluctuations of RH ($\sim 10\%$) occurred in the course of several days following external weather conditions. We assembled colloidal probes by gluing silica spheres of $25.24\mu m$ mean nominal diameter (MicroParticles GmbH $SiO_2$-R-SC93; standard deviation $0.75\mu m$) to rectangular Si cantilevers with molten Shell Epikote resin, using a custom-built micromanipulation stage coupled with an optical microscope [32]. The nominal elastic constant of each cantilever (MikroMasch NSC12/tipless/noAl, type C) was measured by Sader's method [33] and was in the range $k_C{\sim}2.5 - 5.5\mathrm{N/m}$. Softer cantilevers were not used, as they were unable to compensate for the strong adhesion of the pristine silica beads to HOPG.

The applied normal force $F_N$ was estimated as $F_N = \alpha_N \Delta U_N$, where $\Delta U_N$ is the variation of the photodiode current induced by the cantilever vertical deflection and the normal force calibration factor $\alpha_N = k_N^* s_N$. Here, $k_N^*$ is the effective spring constant of the colloidal probe (see [34] and Supplementary Data section S2) and $s_N$ is the vertical deflection sensitivity of the AFM optical lever system. The lateral force calibration factor $\alpha_L$, required to estimate the lateral force $F_L = \alpha_L \Delta U_L$, was determined by means of a diamagnetic levitation spring system [35].

Sliding friction experiments were carried out as reported elsewhere [36,37]. Briefly, to obtain friction force *vs* normal load ($F_f\ vs\ F_N$) characteristics, $F_N$ was decreased every ten lines from a large value $F_N{\sim}700\mathrm{nN}$ to the pull-off value; the corresponding lateral forces, resulting from the difference between forward and backward scans, were averaged on eight lines in between the $F_N$ jump to produce one data point. Ten characteristics acquired over the same region were ensemble averaged, to obtain the average $F_f\ vs\ F_N$ characteristics reported in the graphs [36,37]. Error bars smaller then the symbols size were omitted. To avoid morphological contributions to friction, $F_f\ vs\ F_N$ curves were acquired over atomically-flat areas free of atomic steps, with scan range varying from $11 \times 11\mathrm{nm}^2$ up to $500 \times 500\mathrm{nm}^2$. Friction maps were analyzed with a custom software in National Instruments LabView and they were displayed using WSXM [38].

For the morphological characterization of the colloidal probes, we carried out reverse AFM imaging on a spiked grating (MikroMasch TGT01). We used scan areas of about $2 \times 2\ \mu\mathrm{m}^2$, scan rates $0.5 - 1.0\mathrm{Hz}$ and a normal load $F_N{\sim}50\mathrm{nN}$. Investigation of AFM probe convolution effects was carried out with the aid of the Gwyddion software [39]. Additionally, high resolution SEM imaging of silica beads was completed under a 1kV acceleration voltage using a field-emission SEM instrument (CrossBeam 1540 XB by Zeiss).

**2.3 Modelling nanoscale friction and adhesion by single-asperity contact mechanics.** Within single-asperity nanotribology, the nonlinearity of experimental friction *vs* load curves is traced back to a nonlinear dependence of the tip-sample contact area $A$ on $F_N$, $A = A(F_N)$, thereby assuming a constant interfacial shear strength $\tau$ [40]. The $A = A(F_N)$ relationship is predicted by the continuum Maugis-Dugdale (MD) transition model in its generalized form [41]. This model deals with a sphere-on-flat contact geometry and determines the position of the contact interface along a spectrum of behaviors ranging between two limiting cases, hard contacts with long-range attractive forces or soft contacts with short-ranged adhesion. The former are usually modelled through the simplified Derjaguin-Mueller-Toporov (DMT) theory [42], whereas the latter *via* the Johnson-Kendall-Roberts (JKR) model. In turn, the MD transition model comprehensively treats both limiting and intermediate cases. The location of the contact is embodied by the dimensionless parameter λ which varies from zero to infinity [41]. In practice, for $\lambda < 0.1$ the DMT model applies whereas for $\lambda > 5$ the JKR model is appropriate. Hybrid cases have $0.1 < \lambda < 5$. We obtained estimates of the transition parameter $\lambda$, adhesion energy $\gamma$, interfacial shear strength $\tau$ and contact area at zero load $A_0 = A(F_N = 0)$, by fitting the non-linear portion of $F_f$ $vs$ $F_N$ curves with the analytical solution of the transition model developed by Carpick *et al.* [40]. The Young's modulus and Poisson's ratio used in the fit were respectively 70GPa and 0.2 for the silica spheres, 30GPa and 0.24 for HOPG [43] (see Supplementary Data section S2).

The assumption of a single-asperity contact geometry with curvature radius $R = R_{sphere} \approx 12.6$μm is certainly appropriate to describe the pristine silica spheres in contact with HOPG. However this choice appears questionable when the TL is formed. In this study we provide evidences that the TL does consist of nanosized multilayer graphene flakes protruding from the silica beads' profile, so that a single-asperity contact with effective curvature radius $R \ll R_{sphere}$ reasonably describes the sliding friction between the TL and the atomically-smooth basal plane of graphite.

Regarding adhesion force, we evaluated capillary contributions originated by condensation of ambient water vapor in the AFM probe-sample gap. Following Butt *et al.* [44], we modeled the capillary force for a sphere-on-flat contact geometry and explicitly accounted for the silica bead surface roughness and for TL effects.

**2.4 Atomic-scale friction data: analysis and simulations.** With the aim of possibly analyzing the observed frictional response of the system based on the mechanics of the contact and on the experimental setup, rather than on the specific incommensurability features of the sliding interface, we analyzed experimental atomic-scale friction maps in the framework of the one-dimensional Prandtl-Tomlinson (PT) single-asperity model [45]. In the PT model, the sliding tip is dragged over a sinusoidal potential with amplitude $E_0$ and periodicity $a$. The pulling spring, extended between the position of the tip and the actual position of the AFM moving support, has an effective spring constant $k$ that combines the lateral stiffness of the cantilever with that of the contact. The movement of the tip from one minimum of the potential to the next can be smoothly continuous or abrupt depending on the ratio between the corrugation $E_0$ and the elastic energy, as described by the Tomlinson parameter (or relative corrugation) $\eta$. The $\eta$ parameter is particularly useful, as it signals transitions from dissipative stick-slip motion ($\eta > 1$) to continuous superlubric sliding ($\eta \leq 1$). We estimated $E_0$, $\eta$ and $k$ as a function of normal load $F_N$ according to [45,46]. To this end, for any given value of $F_N$, we first made a careful selection of the lateral force profiles. In fact the actual motion of the AFM probe on HOPG is usually two-dimensional zigzag-type, i.e. at least two prominent jump distances occur when the probe moves along any direction other than the zigzag crystallographic one [47]. Since each jump distance probes the corrugation potential along a specific crystallographic direction, we only considered lateral force traces with periodicity $a \approx 0.21 - 0.25\text{nm}$, corresponding to slip jumps nearly along the zigzag crystallographic direction. Force traces showing different jumps were disregarded. For each selected force profile, the relative corrugation $E_0$, the parameter $\eta$ and the contact stiffness $k$ were evaluated as:

$$E_0 = \frac{aF_{L,max}}{\pi} \qquad (1)$$

$$\eta = \frac{2\pi F_{L,max}}{ak_{exp}} - 1 \qquad (2)$$

$$k = \frac{\eta+1}{\eta} k_{exp} \qquad (3)$$

where $F_{L,max}$ are local force maxima, $k_{exp}$ are the corresponding slopes of the lateral force versus displacement at the beginning of each stick phase, and $a$ are the related jump distances [45]. Only the highest force maxima were considered, in order to limit underestimation of $E_0$ due to thermally-activated jumps [46]. Finally, we ensemble averaged the $E_0$, $k$ and $\eta$ values extracted from several lateral force profiles to obtain mean values for any chosen value of the normal load $F_N$.

Theoretical friction *vs* displacement profiles were obtained as numerical solutions of the one-dimensional Langevin equation for atomic friction:

$$m\ddot{x} + m\gamma x = -\frac{\partial V(x,t)}{\partial x} + \xi(t) \qquad (4)$$

where the total potential $V(x,t)$ is

$$V(x,t) \equiv -\frac{E_0}{2}\cos\left(\frac{2\pi x}{a}\right) + \frac{1}{2}k(vt-x)^2 \qquad (5)$$

Friction force was calculated as:

$$F_L = k(vt-x) \qquad (6)$$

For the effective tip mass we used multiples of $m_0 = 1 \times 10^{-12}$Kg, with the Langevin damping $\gamma = 0.01\text{ns}^{-1}$ aiming at reproducing the experimental force trace profiles. The thermal noise term $\xi(t)$ satisfies the fluctuation-dissipation theorem: $\langle \xi(t)\xi(t') \rangle = 2m\gamma k_B T \delta(t - t')$. We implemented the fourth-order Runge–Kutta algorithm with time step $100\text{ns}$. Temperature T, spring stiffness $k$ and sliding velocity $v$ where chosen to match AFM experiments.

## 3. RESULTS AND DISCUSSION

### 3.1 Signatures for transfer layer formation in nanoscale force spectroscopy

Hereafter, we use $\text{SiO}_2/\text{HOPG}$ to indicate the contact junction between pristine silica beads and HOPG. When a TL is formed, we use $\text{SiO}_2 - \text{TL}$ for the silica beads covered by the TL, and $\text{SiO}_2 - \text{TL}/\text{HOPG}$ for the related contact junctions with graphite. According to background literature [21,24], the formation of an 'ultralow-friction' interface proceeds via triboinduced material transfer from HOPG to the sliding countersurface. We induced material transfer by driving pristine colloidal probes over freshly-cleaved HOPG areas ($2 \times 2\mu\text{m}^2$ up to $8 \times 8\mu\text{m}^2$), with the sliding direction almost perpendicular to the local orientation of the atomic steps. Contact forces between pristine spheres and HOPG were dominated by a large adhesion ($\sim 2 - 3\mu$N); accordingly, the choice of a normal load $F_N$ up to several hundreds of nN had no relevant effect on the TL formation process. The friction maps of Fig. 1a,b attest the typical evolution of the lateral force. Initially, in the absence of TL, lateral force $F_L$ fluctuates around $\sim 100\text{nN}$, and large peaks of $\sim 150 - 300\text{nN}$ localize at the surface step edges as a result of the interaction of the bead with the surface morphology (Fig. 1a). As scanning progresses, random transitions to smaller lateral forces ($\sim 1 - 5\text{nN}$) occur along isolated scan lines (see Supplementary Data section S3), until both the lateral force $F_L$ and the friction loop area undergo a permanent drop over the majority of the scanned area (Fig. 1b). We show below (section 3.2) that such friction drop represents the peculiar signature for the formation of a graphitic TL on the silica bead. We anticipate that a simultaneous drop of contact adhesion $F_A$ takes place, too.

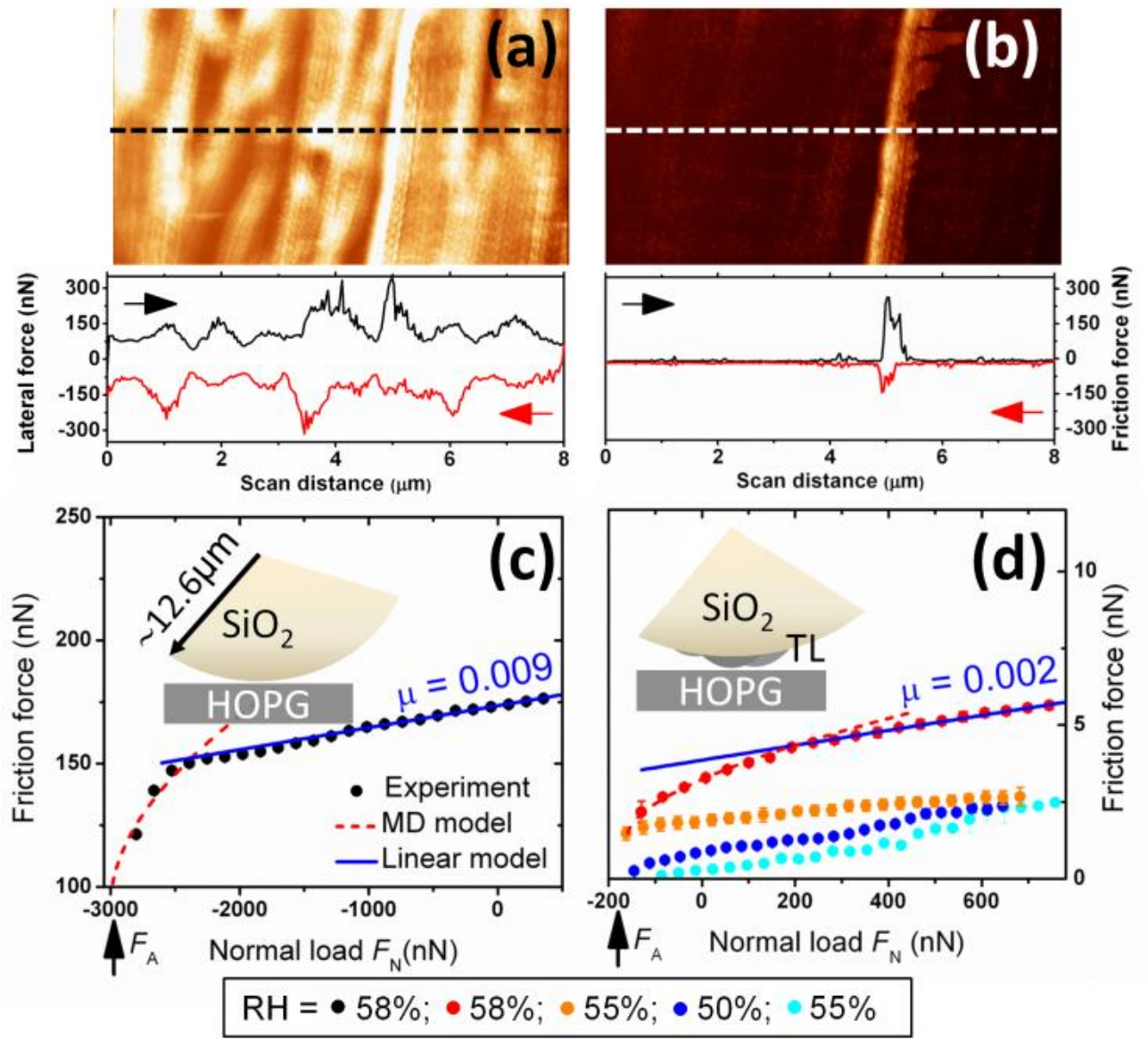

**Fig. 1.** (a) Representative friction map and related friction loop for a pristine silica bead sliding on HOPG ($SiO_2$/HOPG contact; $F_N = 350\text{nN}$). (b) Friction map and friction loop over the same region probed in (a), after formation of a triboinduced TL ($SiO_2$-TL/ HOPG contact). (c) Typical load-dependent friction characteristic for the $SiO_2$/HOPG contact. (d) Several load-dependent friction characteristics for the $SiO_2$-TL/HOPG contact (graph legend as in (c)). RH is specified for each friction *vs* load curve.

To gain a deeper insight we systematically performed variable-load friction measurements over atomically-flat areas of graphite (from $200 \times 200\text{nm}^2$ to $500 \times 500\text{nm}^2$), respectively before and after the formation of the TL. Fig. 1c shows a representative $F_f$ $vs$ $F_N$ curve for the $\text{SiO}_2/\text{HOPG}$ contact. A large adhesion $|F_A| \approx 3.0\mu\text{N}$ dominates the contact. Furthermore, a linear regime progressively evolves into a non-linear one as $F_N$ approaches the pull-off point $F_N \sim F_A$. In the linear regime, the friction force is $F_f \sim 150 - 170\text{nN}$ and linear regression gives a friction coefficient $\mu = 0.009$. Interpolation of the non-linear portion with the single-asperity MD model gives $\lambda \sim 1.7$, hence the contact is closer to the JKR limit. Using the nominal curvature radius $R = R_{Sphere} \approx 12.6\mu m$, one gets the effective adhesion energy $\gamma \approx 60\ \text{mJ}/\text{m}^2$, the shear strength $\tau \approx 2.2\text{MPa}$ and the zero-load contact area $A_0 \sim 9 \times 10^4\text{nm}^2$. This area corresponds to a circular contact with mesoscopic diameter $\sim 340\text{nm}$.

Albeit the formation of the TL was characterized by a robust friction drop (Fig. 1a,b), we found great variability in the $F_f$ $vs$ $F_N$ curves describing the 'ultralow-friction' $SiO_2 - TL/HOPG$ interface. In fact, upon the formation of the TL, the response of the colloidal probe was reproducible if interrogated over atomically-flat areas. However, random transitions from one characteristic to another often occurred for the same probe, as a result of its interaction with the graphite surface steps. This indicates that the TL was prone to structural rearrangements (see also section 3.4). Fig.1d reports a selected ensemble of 'ultralow-friction' $F_f$ $vs$ $F_N$ characteristics. Compared to Fig. 1c, the friction force $F_f$ is reduced by a factor $\sim 30$ to $\sim 70$ and linear regression gives $\mu = 0.001 - 0.002$. Such values agree with those reported for other superlubric mesoscale contacts on graphite [48]. Additionally, the adhesion force is greatly suppressed, $|F_A| \approx 100 - 200$nN. Remarkably, most of the characteristics show a non-linear trend for tensile loads. From interpolation with MD model, one obtains $\lambda \geq 1.6$ (JKR limit). Extraction of interfacial parameters is however not trivial due to the presence of the TL. Choosing a contact radius $R = R_{Sphere}$ would provide extremely low values for interfacial adhesion ($\gamma \sim 3\text{mJ/m}^2$) and shear stress ($\tau \sim 0.17$MPa). This choice appears questionable in view of the TL morphology, as studied by reverse AFM. In section 3.2 we show that the TL is rough on the nanometers scale and the bead contact with the atomically-smooth basal plane of graphite can be ascribed to an individual nanocontact, i.e. the highest triboinduced nanoasperity [49]. It is thus more appropriate to treat the 'ultralow-friction' $SiO_2 - TL/HOPG$ interface as a single-asperity contact with curvature radius $R \ll R_{sphere}$. Specifically, the choice $R = (300 \pm 100)$nm appears reasonable with respect to the range of $R$ values ($\sim 200 - 400$nm) obtained from reverse AFM images of several triboinduced nanoasperities. With this assumption one obtains $\gamma \sim 130\text{mJ/m}^2$, $\tau \sim 2$MPa and $A_0 \sim 1.0 \times 10^3 \text{nm}^2$. Table 1 summarizes the contact parameters extracted from several $F_f$ $vs$ $F_N$ curves associated to different probes.

| | $\gamma$ $\mathrm{mJ/m^2}$ | $\tau$ MPa | $A_0$ $\times 10^3\mathrm{nm}^2$ | $\mu$ $\times 10^{-3}$ |
|---|---|---|---|---|
| **$SiO_2$/ HOPG** *R* = 12.6μm ± 0.75μm | $50 \pm 20$ | $1.9 \pm 0.7$ | $70 \pm 30$ | $9 \pm 2$ |
| **$SiO_2$ – TL / HOPG** *R* = 300nm ± 100nm | $130 \pm 50$ | $2.0 \pm 1.0$ | $1.03 \pm 0.13$ | $1.5 \pm 0.5$ |

**Table 1.** Contact parameters extracted from experimental $F_f$ $vs$ $F_N$ curves (see text). Variability ranges reflect interfacial heterogeneity for different probes and uncertainty on the curvature radius $R$. The role of the TL stems out clearly through the decrease of $A_0$ and $\mu$.

According to Table 1, the main effect of the TL is to reduce the real contact area $A_0$ by more than one order of magnitude, that turns the mesoscopic $\mathrm{SiO_2/HOPG}$ contact into a nanoscopic one. As the shear strength $\tau$ does not vary substantially, the MD model ascribes the breakdown of the friction force $F_f$ mainly to the decrease of $A_0$. Interestingly, $\tau \sim 1 - 2\mathrm{MPa}$ is an order of magnitude smaller that the shear stress measured in AFM experiments using conventional nanosized tips [36]. An explanation for the decrease of the adhesion force $F_A$ (from $\sim 2 - 3\mu\mathrm{N}$ to less than $200\mathrm{nN}$) needs additional insight into its origin. It is known that capillary adhesion, due to a water meniscus condensed from ambient air at the probe-HOPG contact, dominates over van der Waals (vdW) forces whenever hydrophilic surfaces are in contact under ambient conditions [50]. Calculations indicate that the capillary force between the pristine silica microsphere and HOPG amounts to $F_{cap} \sim (2.6 \pm 0.4)\mu\mathrm{N}$ at a relative humidity $\mathrm{RH} \sim 60\%$ (see Supplementary Data section S4). Thus $F_{cap}$ is expected to be the main contribution to the measured adhesion force $F_A$ at the $SiO_2$/HOPG interface. In section 3.2 we show that the formation of the TL implies a local increase of the beads surface roughness from tenths of to a few nanometers. As capillary force greatly decreases when roughness overcomes the characteristic length scale of capillary condensation, i.e. the Kelvin length $\lambda_K = 0.52\mathrm{nm}$, one concludes that the main effect of the TL is to suppress capillary adhesion between the micrometric bead and graphite. At the $SiO_2 - TL/HOPG$ interface, adhesion $F_A$ is thus mostly governed by attractive forces within the graphitic, nanosized contact area $A_0$. Notably, the MD model predicts $\gamma \sim 130\ mJ/m^2$, not far from the $160 - 220 mJ/m^2$ measured in conventional single-asperity AFM experiments on graphite [36,51].

### 3.2 Morphology and contact mechanics of the triboinduced transfer layer

We routinely performed reverse AFM imaging to characterize variations of the surface morphology of the silica beads, taking place upon the TL formation. Results for a representative colloidal probe are reported in Fig. 2a-f. In the pristine state, the bead's surface is homogeneous and free from localized bumps (Fig. 2a); cross sections are satisfyingly fitted by circular fits, with a curvature radius close to the nominal one (Fig. 2b). Surface roughness, evaluated after subtraction of the spherical shape, amounts to $\sigma \sim 0.5 nm$ over surface areas of $\sim 1 \mu m^2$. Inspection of the same probe after reduction of both $F_A$ and $F_f$ reveals the presence of a TL. As shown in Figs. 2c-f, this consists of flakes with lateral size $\sim 50 nm$, clustered within an apical region of the silica bead. This region corresponds to the contact region with HOPG. Cross sectional heights show that the flakes protrude a few nanometers away from the bead's circular profile ($\sim 2 - 5 nm$), in fact representing isolated asperities spaced tens of nanometers apart. The TL thus gives an overall increase of the bead's surface roughness within the contact region, from $\sim 0.5 nm$ to about $\sim 2 nm$. Importantly, this value agrees with the thickness of nanoflakes directly imaged by conventional AFM onto worn HOPG regions, after the TL formation (see Supplementary Data section S3). It also nicely agrees with the $2.3 nm$ –thick TL measured by Li *et al.* by means of high-resolution transmission electron microscopy (HRTEM), for the case of AFM friction experiments similar to ours [24]. Additionally, we confirm *via* reverse friction maps the ultralow friction response of such flakes compared to silica, this being an indication of their graphitic nature. This goes in par with the evidence that lubricity comparable to bulk graphite has been reported also for smaller flakes (lateral size $\leq 30 nm$), tribotransferred at the apex of AFM tips [52] or supported on technical substrates [53]. Hence the bead surface is covered by superlubric nanoasperities. This is indeed a general conclusion, that we verified by reverse AFM imaging of different colloidal probes. Figure 2g further shows the excellent agreement between the TL topography as resolved by reverse AFM imaging, and results from high-resolution SEM (see Supplementary Data section S2 for details on contact spot location in SEM).

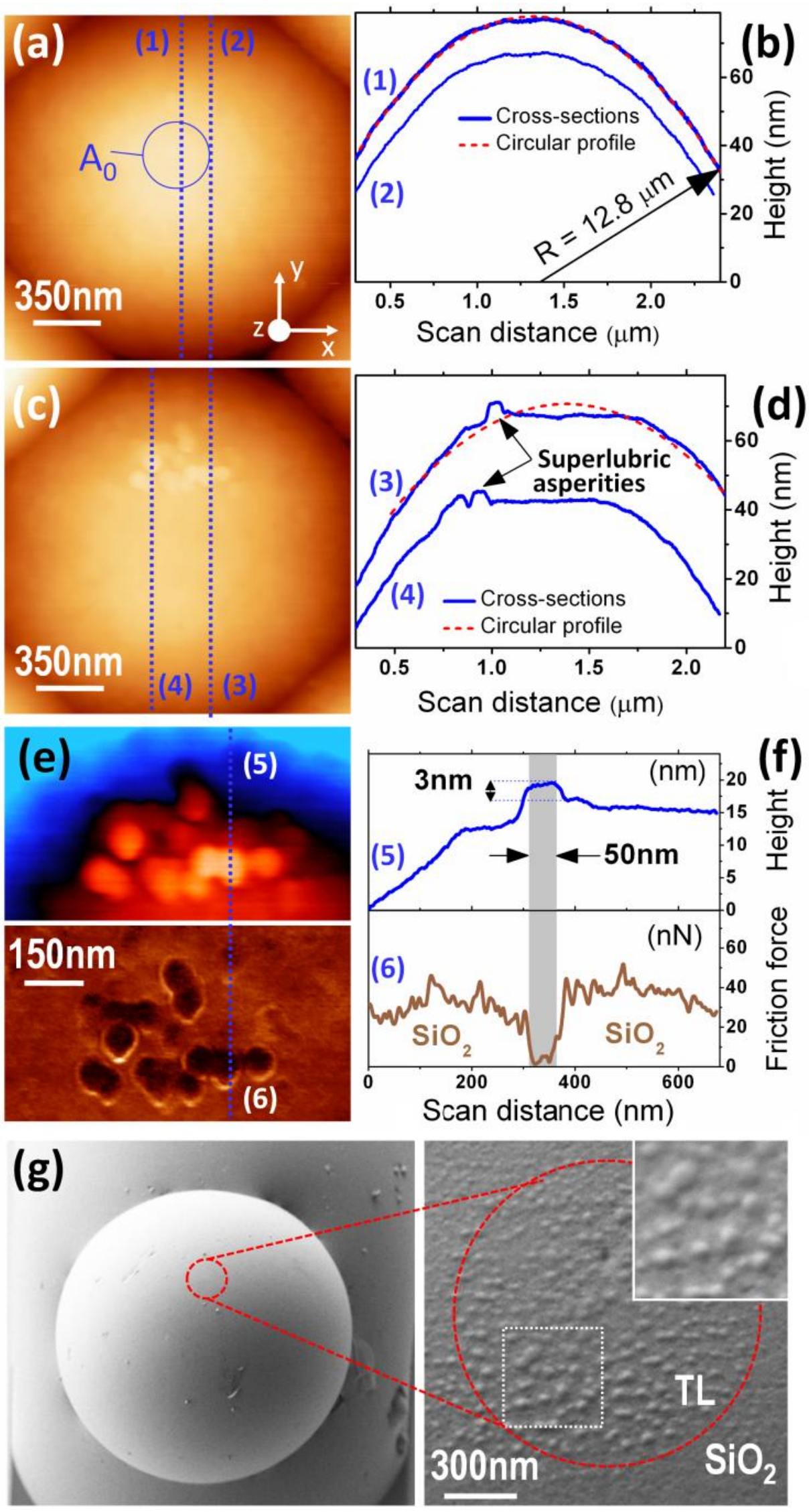

**Fig. 2.** (a) AFM topography of a pristine silica bead, with $A_0 = 7 \times 10^4 \mathrm{nm}^2$ highlighted for clarity (see Tab.1). (b) Cross sections along the dashed lines in (a) show the smooth circular profile of the bead. (c) Topography of the same bead after formation of the TL. (d) Cross sections along the dashed lines in (c) show that the TL consists of several protruding nanoasperities. (e) Magnified view of the TL nanoasperities in (c) reveals the presence of multilayer graphene flakes (topography and friction in the top and bottom panels respectively). (f) Cross section along the dashed lines in (e) confirm the ultralow friction response of the flakes compared to silica. (g) SEM micrographs of a silica bead (not the same in a-f) and of the contact area, together with a high-resolution magnification (inset) of the tribotransferred flakes. The flakes appear only within the contact region, and not on outer surface portions.

We point out that the occurrence of several superlubric asperities on the beads' surface does not contradict the choice to interpolate 'ultralow-friction' $F_f$ $vs$ $F_N$ curves by means of the single-asperity MD model. To clarify the role played by the triboinduced nanoasperities in contact mechanics we observe that the formation of the TL was accompanied – besides the reduction of the contact forces $F_A$ and $F_f$ – by appearance of multiple-tip effects (see also Supplementary Data section S5). Qualitatively, such effects arise in AFM imaging whenever HOPG has local roughness (e.g. due to multilayer steps or ambient contaminants) with aspect ratio comparable to the surface roughness of the $\mathrm{SiO_2 - TL}$ probe [54]. In such cases, AFM maps display repetitive patterns that reflect the spatial arrangement of the superlubric asperities involved in contact mechanics. On the other hand, when the same probe slides over the atomically-flat basal plane of graphite, normal and shear forces are likely supported by one asperity only corresponding to the most prominent one. Therefore the 'ultralow-friction' $F_f$ $vs$ $F_N$ characteristics of Fig. 1 can be properly analyzed by the single-asperity MD model. Given the $\sim 50\mathrm{nm}$ typical lateral size of a flake, this argument implies a single-asperity contact area $\leq 50^2\mathrm{nm}^2 = 2.5 \times 10^3\mathrm{nm}^2$ at the $\mathrm{SiO_2 - TL/HOPG}$ junction. Such area is only a factor 2 higher than $A_0 \sim 1.03 \times 10^3\mathrm{nm}^2$ reported in Tab.1.

To further support previous discussion on a quantitative base, we recall that morphological dilation allows to predicts the multiple-tip patterns induced by a given AFM probe [54]. In the present case it can be exploited to identify the superlubric nanoasperities involved in contact mechanics for different values of graphite roughness. As a proof of concept, we focus hereafter on the colloidal probe morphology depicted in Fig. 2c.

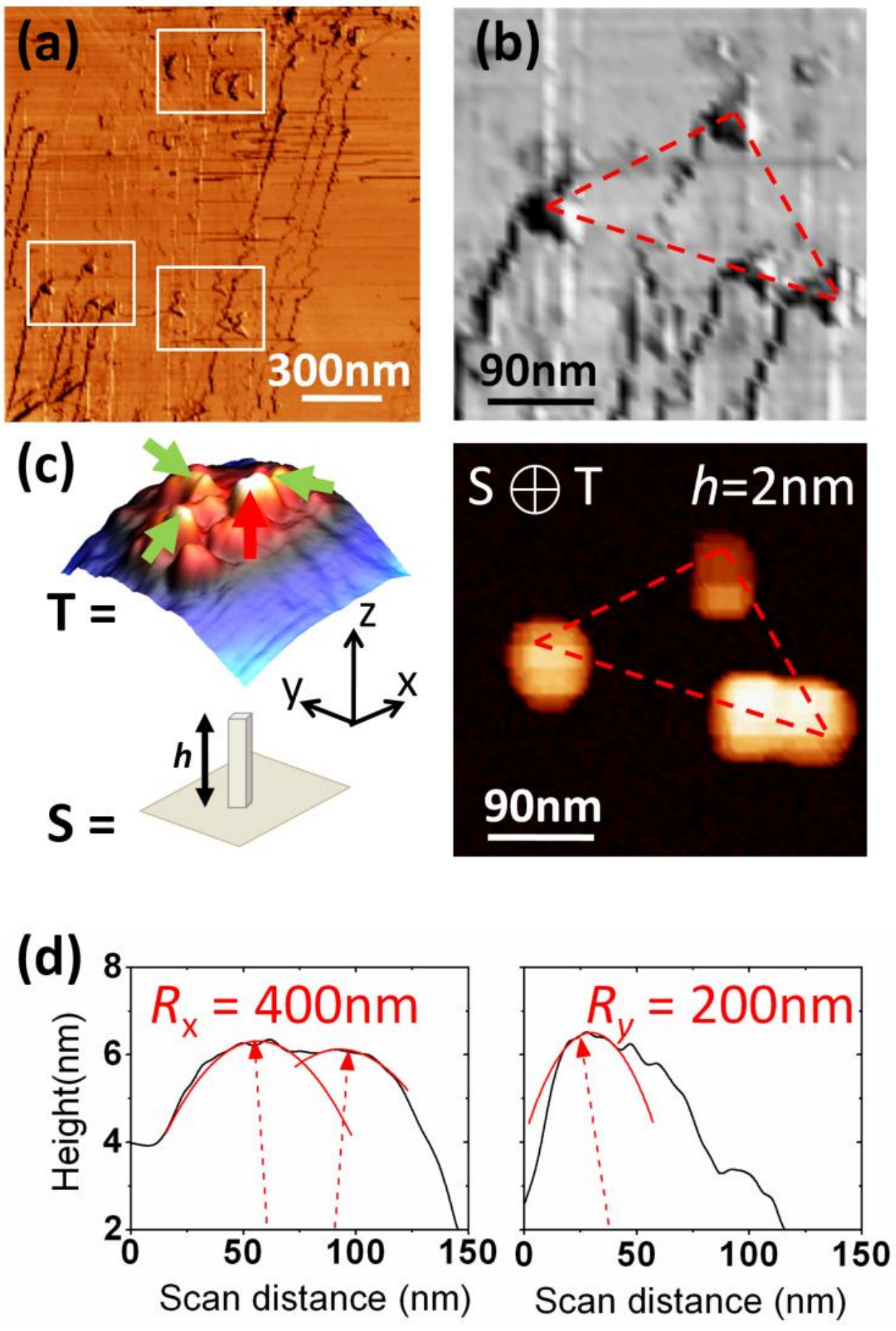

**Fig. 3** (a) Friction map acquired on HOPG by means of the colloidal probe in Fig. 2c. A multiple-tip pattern (highlighted by rectangles) repeats on the map. (b) Magnification of the pattern in (a), showing a triangular shape with four main spots. (c) The multiple-tip pattern in (b) is theoretically predicted by morphological dilation ⊕ between a model surface spike S, and the surface morphology T taken from Fig. 2c. The four triboinduced nanoasperities contributing to the pattern generation are indicated by arrows. (d) Cross-sectional heights and curvature radii for the highest nanoasperity (indicated by the red arrow in (c); x an y directions as in (c)).

Figures 3a,b show a typical friction map acquired with that $\mathrm{SiO_2-TL}$ probe on a stepped region of HOPG. A triangular-shaped pattern with four main spots repeats over different portions of the map, due to interaction of a few superlubric asperities with the local surface roughness. Using the morphology of Fig. 2c, we have implemented the dilation of a square based surface spike mimicking localized graphite roughness (Fig. 3c and Appendix B). For a spike height in the range $h = 1 - 5$nm, the dilated topography shows a robust triangular-shaped pattern that excellently agrees with that observed experimentally. Four superlubric asperities contribute to generate such triangular pattern (green and red

arrows in Fig. 3c). More importantly, however, the pattern disappears in favor of one individual spot for spike heights $h \leq 0.1\text{nm}$ (Appendix B, Fig. B.3f). This confirms that only one superlubric asperity was in contact with HOPG when the colloidal probe slid over atomically-smooth graphite. Cross sectional heights for such tip give curvature radii $R_x \sim 400\text{nm}$ and $R_y \sim 200\text{nm}$ along two orthogonal directions (see the three orthogonal axis of Fig. 2c). This information is highly relevant to study the contact mechanics of the TL, as the effective parameter $R$ enters the single-asperity MD analysis (through Eqs. (4),(5)).

In line with such evidences, we repeatedly carried out cross-sectional analysis of the protruding superlubric asperities for different silica probes, and found $R \sim (300 \pm 100)\text{nm}$ to be a reasonable choice for the average curvature radius at the $\text{SiO}_2 - \text{TL/HOPG}$ interface. Results of the MD analysis with such $R$ value are resumed in Tab. 1.

### 3.3 Atomic-scale dissipation with the triboinduced transfer layer

We addressed the elementary dissipation mechanisms at the $\text{SiO}_2 - \text{TL/HOPG}$ interface by analyzing lattice-resolved lateral force maps (scan range $11 \times 11\text{nm}^2$). Within a broad range of normal loads, $100\text{nN} \lesssim F_N \lesssim 400\text{nN}$, such maps showed stick-slip motion with the hexagonal symmetry of the graphite lattice (Fig. 4a). Fast Fourier Transform (FFT) analysis attested small variations of the in-plane orientation of the graphite lattice in the course of the experiments, the probe sliding direction being usually slightly off ($\leq 10°$) the armchair crystallographic direction (see inset of Fig. 4a). Despite the measurement of extremely low friction coefficients at the $\text{SiO}_2 - \text{TL/HOPG}$ (Tab.1), the ubiquitous presence here of a stick-slip motion, however small, signals the appearance of an interlocking mechanism between the TL nanoasperity of the tip and the underneath graphitic surface. As discussed later on, the ideal suppression of all the energy barriers to sliding motion due to cancellation of substrate lateral forces - as prescribed by structural superlubricity - may be hampered within such range of normal loads by the concomitant occurrence of different pinning effects.

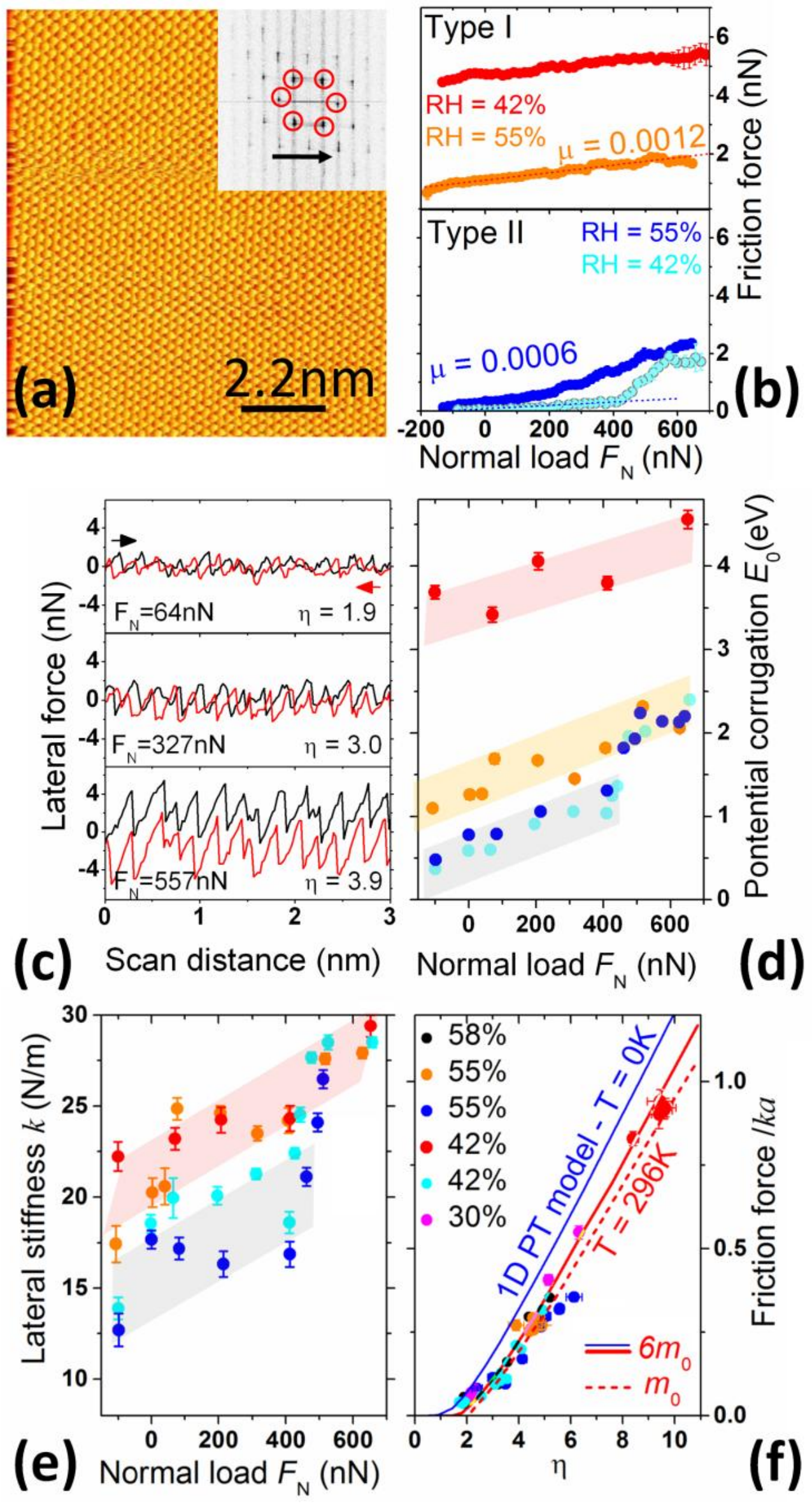

**Fig. 4** (a) Atomic-scale friction map recorded after the TL formation ($F_N = 198\text{nN}$). Inset: main spots of the 2D FFT and fast sliding direction (black arrow). (b) Representative friction *vs* load curves ($v = 33\text{nm/s}$). (c) Load-dependent friction loops for the lowest friction curve in (b). (d,e) Load-dependent potential corrugation $E_0$ and contact stiffness $k$ extracted from the friction loops of curves in (b). (f) Normalized friction force $F_f^*$ *vs* $\eta$ and comparison with the 1D PT model ($m_0 = 1 \times 10^{-12}\text{Kg}$; see text). RH is specified for each friction *vs* load curve.

As for the AFM experiments conducted over longer length scales ($200 - 500\text{nm}$, Fig. 1d), we found great variability of $F_f$ *vs* $F_N$ curves also at the atomic scale. Some representative characteristics, obtained with the same colloidal probe, are reported in Fig. 4b. Their appearance reflects changes in the fine details of the $SiO_2 - TL/HOPG$ contact, taking place in the course of the experiments and regardless of the RH value. On one hand there are curves showing a weak dependence on load and finite friction force $F_f \geq 1\text{nN}$ down the pull-off point ('type I' friction curves). On the other side, there are entirely different characteristics showing vanishing friction at the lowest loads and a pronounced slope increase at some point ('type II' friction curves). Notably, for one of them, an ultralow friction force $F_f \leq 0.3\text{nN}$ persists up to $F_N \sim 400\text{nN}$. With a pull-off force $|F_{adh}| \sim 80\text{nN}$ and a friction coefficient $6 \times 10^{-4}$, such $F_f\ vs\ F_N$ curve perfectly matches results previously discussed in the framework of incommensurate contacts [24]. Inspection of the related friction loops (Fig. 4c) shows that mechanical dissipation greatly evolves with the normal load $F_N$, from nearly-frictionless continuous sliding ($F_N = 64\text{nN}$) and regular stick-slip ($F_N = 327\text{nN}$), to highly-dissipative (multi-slip) stick-slip ($F_N = 557\text{nN}$). This trend agrees very well – both qualitatively and quantitatively – with atomic-friction experiments conducted on graphite by means of sharp AFM tips [55,56], and suggests to exploit the single-asperity PT model to rationalize the friction behavior of the $\text{SiO}_2 - \text{TL/HOPG}$ contact. To this end, we estimated the contact parameters $E_0$, $k$ and $\eta$ of the PT model as a function of $F_N$. For comparative purposes, analysis was carried out for each contact behavior (i.e. $F_f\ vs\ F_N$ characteristic) of Fig. 4b. Results are summarized in Fig. 4d-f. Specifically, the potential corrugation $E_0$ increases with $F_N$ (Fig. 4d) and assumes the smallest values (0.37eV – 1.0eV) for 'type II' curves, whereas higher barriers (1.0 – 2.0eV, 3.2eV – 4.2eV) are associated to 'type I' characteristics. Similar trends regard the effective contact stiffness $k$. This analysis allows mapping different $F_f\ vs\ F_N$ curves into the adimensional scaling $F_f^*\ vs\ \eta$, where $F_f^* \equiv F_f/ak$ is the normalized friction force. In this way it turns out that all the experimental $F_f\ vs\ F_N$ characteristics (acquired in the broad range $\text{RH} \sim 30\% - 60\%$) follow reasonably well the thermally-activated 1D PT model (Fig. 4f). Here, model predictions rely on numerical solutions of the Langevin equation for atomic friction, with temperature $\text{T} = 296\text{K}$ and sliding velocity $v = 30\text{nm/s}$ closely matching experiments (see also Supplementary Data sections S6, S7). Notably, thermal energy assists the contact asperity of the TL to jump from one minimum of the potential to the next, which gives a friction force $F_f^*$ systematically smaller (thermolubricity) than that predicted by the athermal ($\text{T} = 0\text{K}$) PT model [46,57]. From previous analysis we are led to conclude that small variations of the $\text{SiO}_2 - \text{TL/HOPG}$ contact make the system to populate different portions of the $F_f^*$ $vs$ $\eta$ curve. This is clear from clustering of each friction $F_f$ $vs$ $F_N$ characteristic into a specific region along the $\eta$ axis. In particular the 'type II' curves mostly populate the small dissipation region $1.7 \leq \eta \leq 4$. Here, the $SiO_2 - TL/HOPG$ contact transitions from continuous (thermolubric) sliding to stick-slip motion. Within such region, any increase of $F_N$ gives an increase of the energy barrier $E_0$ that needs to be surmounted for relative lateral motion. The contact stiffness $k$ increases too, but such variation is more than compensated for by the increased corrugation of the potential. The Tomlinson parameter $\eta \propto E_0/k$ thus grows with $F_N$, causing the transition from continuous sliding to a stick-slip mode [56]. On the other hand, curves of 'type I' populate the region $\eta > 4$, where the $SiO_2 - TL/HOPG$ contact dissipates mechanical energy only by stick-slip instabilities [56]. For completeness we underline that we also found a logarithmic increase of the friction force with sliding velocity, with the occurrence of stick-slip instabilities (see Supplementary Data sections S6).

We observe that strong similarities exist between the atomic friction phenomenology depicted in Fig. 4 and background literature from single-asperity nanotribology. Dienwiebel *et al.* [58] already recognized two entirely different types of friction behaviors - named respectively 'high-friction state' and 'low friction state' - for a nanosized flake dragged by a sharp AFM tip on HOPG. They also noticed that $F_f$ $vs$ $F_N$ curves switched randomly back and forth between the two types. This was demonstrated to originate from the difference in orientational misalignment between the flake and the graphite substrate, the 'high friction state' corresponding to a fully commensurate contact and the 'low friction state' to an incommensurate one. Friction duality, with two distinct regimes of finite and vanishing friction, was also observed in nanoparticles manipulation experiments conducted on HOPG [59]. Again, the vanishing friction state was ascribed to interfacial lattice mismatch at the contact interface, whereas the finite friction state was attributed to some pinning effect, e.g. interfacial defects or ambient contaminants. Accordingly, we speculate that the 'type II' regime of Fig. 4b very likely reflects a condition of structural lubricity, dictated by in-plane misalignment between an individual triboinduced flake and graphite, holding up to the threshold load $F_N \sim 100\text{nN}$ (where stick-slip sets in). Such onset is substantially larger than the $\sim 1 - 40\text{nN}$ thresholds documented for atomic and molecular contacts [45,55,60,61]. Also, the potential corrugation $E_0$ and contact stiffness $k$ in the superlubric state ($E_0 \sim 0.3 - 0.6 \mathrm{eV}$, $k \sim 12 - 15 \mathrm{N/m}$) overcome those found for atomic contacts on graphite and graphene ($E_0 \sim 0.08 - 0.4 \mathrm{eV}$, $k \sim 1 - 5 \mathrm{N/m}$ [58,60,62]. All together, these facts likely reflect the increased lateral size of the $\mathrm{SiO_2 - TL/HOPG}$ contact compared to the case of a sharp AFM probe. In line with the above reasoning, we attribute the load-driven transition from continuous superlubric sliding to stick-slip motion in 'type II' curves (around $F_N \sim 100 \mathrm{nN}$) to the progressive emergence of contact pinning effects. One possibility relies on the increase of the degree of interfacial commensurability, e.g. due to small (reversible) in-plane rotations of the tribotransferred flake bearing contact [46]. In fact, Jinesh *et al.* [46] proved that variations in the misfit angle between an individual tribotransferred flake and graphite make $\eta$ to span the range $2 < \eta < 7$. Besides such observations, Dienwiebel *et al.* [5] and Wijk *et al.* [63] demonstrated that for an incommensurately stacked flake, a sudden and reversible increase of friction takes place at a sufficiently large compressive load $F_N$ due to pinning effects at the flake contact boundary [63,64]. Such effects might explain both the sudden rise of friction for the ultralow 'type II' $F_f \ vs \ F_N$ curve of Fig. 5b (at $\sim 400 \mathrm{nN}$), as well as the increase of friction in the $\sim 200 - 400 \mathrm{nN}$ range for all 'type II' characteristics (see Supplementary Data sections S6). Following discussion above, we ascribe 'type I' regime to the single-asperity contact fully dominated by pinning effects (i.e. increased interface commensurability and/or contact edge effects). Under such conditions we find the highest interfacial corrugation and contact stiffness ($E_0 > 1 \mathrm{eV}$, $k \sim 15 - 25 \mathrm{N/m}$), remarkably not far from those reported for large-area graphite flakes dragged by AFM over HOPG in a commensurate state ($E_0 \sim 1 \mathrm{V}$, $k \sim 20 - 40 \mathrm{N/m}$ [65]). Materials characterization is consistent with interfacial registry arguments and contact edge effects. In fact, Li *et al.* [24] confirmed by HRTEM the layered structure of FLG flakes tribotransferred at the silica/HOPG interface. However, as the interlayer spacing of 0.38nm is slightly larger than that of pristine graphite (0.33nm), there might be the introduction of oxygen-containing functional groups during the tribo-exfoliation of the graphite nanosheets. Hence, ambient contaminants and sliding-induced defects might contribute to pinning effects [66]. A richer information on flakes (carbon hybridization, defects, adventitious contaminants) might come from investigations exploiting different spectro-microscopy techniques [67,68], albeit this appears challenging due to flakes size (~50nm) and sparse coverage at the ~300nm contact interface.

### 3.4 Friction variability with the triboinduced transfer layer: limitations and opportunities

We argue that the random character of both interfacial wear and tribotransfer processes are at the base of the friction variability exhibited by the $SiO_2 - TL/HOPG$ contacts (see Fig. 1d, 4b). In Fig. 5a we suggest four main steps providing a comprehensive explanation for the evolution of the TL in the course of friction experiments. The starting point is the contact mechanics between the pristine silica bead and freshly cleaved HOPG (step I). Experiments indicate that such a contact is dominated by a large adhesion, $F_A \sim 2 - 3\mu N$ (see Fig. 1). Data interpolation with the MD model predicts a zero-load contact area $A_0 \sim 7 \times 10^4 nm^2$ (Tab.1), that corresponds to a contact diameter of $\sim 300nm$. Accordingly, the nominal contact pressure at the pristine $SiO_2/HOPG$ interface amounts to $\sim F_A/A_0 \approx 40MPa$. Although this stress level is insufficient to initiate wear within the graphite basal plane (in-plane scratching failure is far above $\sim 1GPa$ [69,70]), the surface roughness superimposed on the pristine silica beads ($\lesssim 1nm$) can substantially increase the local contact pressure, and initiate adhesive wear and peel-induced fracture at the surface-exposed step edges. This scenario is also invoked to explain graphene wear in microscale tribotests under relatively small contact pressures [71].

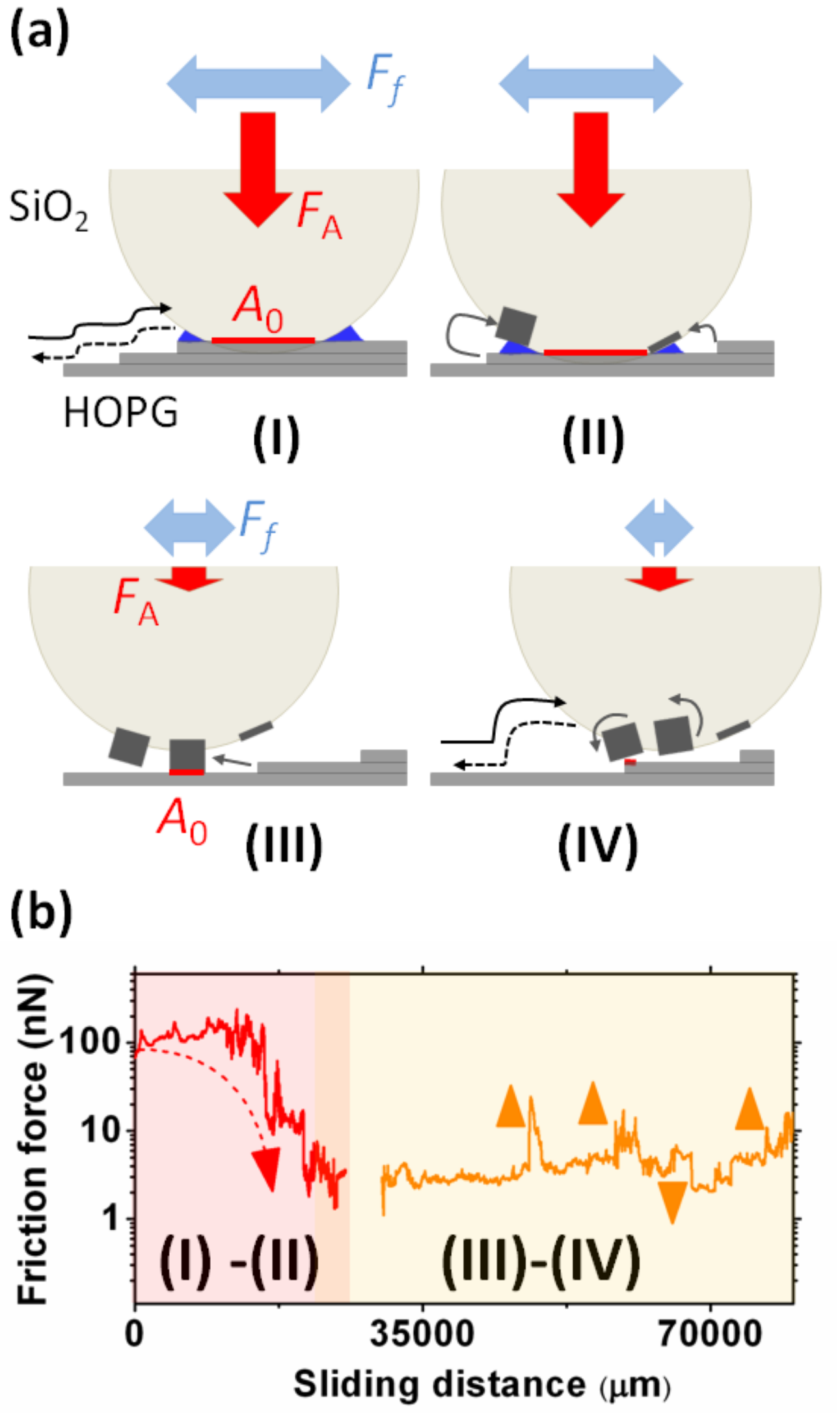

**Fig. 5.** (a) Schematics with four main steps, I to IV, governing the formation and structural rearrangements of the TL (see text). (b) Friction force *vs* sliding distance. Friction breakdown (red line and arrow) and friction fluctuations (orange line and triangles) mark the evolution of the TL. Tentative positions for steps I-II and III-IV are reported along the friction trace.

Nanoscale asperities are in fact known to initiate wear of graphite step edges at a critical normal load about two orders of magnitude lower than that in the interior regions [69,70]. Step edge fracture and peeling are routinely reported for $F_N \sim 100 - 300\text{nN}$ and $F_L \sim 50 - 100\text{nN}$ [69,72]. Hence wear of step edges is expected to be highly active under the typical sliding conditions of the pristine beads ($F_N \sim F_A \sim 2 - 3\mu\text{N}$, $F_L > 100\text{nN}$). The TL formation then proceeds via tribotransfer of flakes from HOPG to the silica bead, followed by their effective incorporation within the contact area. This process however depends on the adhesion and friction properties of each flake to the silica surface. On one hand (step II), we speculate that flakes loosely attached to silica are pushed away by friction forces from the centre to the periphery of the contact, or they are even released back to graphite. On the other side (step III), a flake remains within the contact area only if firmly attached to the silica surface. As soon as this happens, $A_0$ decreases by a factor $> 10$ (Tab. 1). Also adhesion and friction forces decrease ($F_A \sim 200\text{nN}$, $F_L < 10\text{nN}$), which in turn hampers any further abrasive wear at the graphite surface steps. In this way, the contact interface remains temporarily locked into a particular configuration of tribotransferred nanoasperities. As for the pristine $SiO_2/HOPG$ state $F_L \sim 100 - 300\text{nN}$, one might speculate that a stable TL mostly involves flakes that sustain such large shear forces by means of a high flake/silica static friction. Both the systematic observation of multiple-tip effects in AFM imaging (Fig. 3), and the ubiquitous single-asperity PT phenomenology at atomic scale (Fig. 4), indicate that the friction response of mesoscopic silica-graphite junctions depends in such case on the specific energy landscape experienced by the topographically-highest, triboinduced nanoasperity. In view of the typical lateral size of the tribotransferred flakes ($\sim 50\text{nm}$), it is likely that such an asperity involves only one flake. However, we cannot exclude the involvement of a few neighbour flakes. In fact, friction forces experienced by such flakes should sum up to give the total friction measured by AFM. Even if the relative orientation of such flakes is random one with respect to the other, it is unlike to have full cancellation of their lateral force contributions because the degree of orientational disorder for a few-flakes contact does not fit the high degree of disorder required for the manifestation of structural lubricity (e.g. see [73]). Therefore, we do not expect qualitative differences from the case of a single-asperity (single-flake) nanocontact. According to AFM experiments exploiting graphene-wrapped probes [52], an individual nanosized FLG flake with thickness $\sim 2 - 15\text{nm}$ has load-bearing capacity largely exceeding $\sim 1\mu\text{N}$ and can provide ultralow friction coefficients $\mu \sim 10^{-4} - 10^{-3}$ when sliding against bulk graphite. As in the present case $F_N \leq 700\text{nN}$, we are well within the limits explored by such studies and fully consistent with findings thereof. Finally (step IV), reorganizations of the triboinduced asperities are possible, triggered by random interactions of the bead with the HOPG atomic roughness. We believe that such interfacial rearrangements are at the base of the friction fluctuations and variability of experimental friction *vs* load curves. In Fig. 5b, we propose a closer correspondence of steps I to IV with the friction trace displaying the typical evolution of the bead-graphite contact with the sliding distance. After formation of the TL (before $\sim 3 \times 10^4 \mu\text{m}$), well-separated random transitions of the friction force appear superimposed on a low base level $F_L \sim 3\text{nN}$. Random jumps of $F_L$ are extensively documented in micro- and macro-tribological experiments on HOPG [17,71]. Here, the AFM colloidal probe crosses the surface exposed step-edges many times (scan area $50 \times 50 \mu\text{m}^2$) and at high sliding velocity ($v = 15 \mu\text{m/s}$) (see Fig. 1,4 for friction response on atomically-flat areas at $v < 1 \mu\text{m/s}$). Friction transitions indicate dramatic variations of the contact junction, with rise of $F_L$ up to $\sim 20\text{nN}$. This confirms that atomic steps provide a relevant source of perturbation to the low-friction state [17,71]. Interestingly, over such a long distance $\sim 4 \times 10^4 \mu\text{m}$, we never observe full recovery of the high-friction state of the pristine $\text{SiO}_2/\text{HOPG}$ interface. Hence, once the flakes are trapped at the silica-HOPG contact, they are not easily expelled from there (for the loads $< 1\mu\text{N}$ typically applied in this study). Flakes removal from the contact area might probably take place if the $\text{SiO}_2 - \text{TL/HOPG}$ tribosystem was pushed (by normal load and sliding velocity) to achieve shear stresses comparable or higher than in the pristine state (which was not our scope here). As graphite atomic roughness represents a clear limitation for stable superlubricity, we foresee the possibility to mitigate its effect by improving adhesion of tribo-transferred flakes (e.g. by silica surface treatments [74]) or by reducing the surface density of atomic steps with tailored graphitic substrates (e.g. *via* polymer-free wet-transfer of CVD graphene on technical substrates [75]).

The most direct implication of our results is to offer a relatively new framework to understand ultralow friction states for graphene-functionalized microprobes [24,48,76]. As crystalline flakes-wrapped AFM probes are rapidly evolving to include other 2D materials [23,29,77], and structural superlubricity was reported in some cases [52], our results might be of relevance also in such context. In detail it seems promising to investigate the generality of the previous picture with other 2D lubricants, such as Hexagonal Boron Nitride (h-BN) and Transition-Metal Dichalcogenides (TMDs). Crystalline nanosheets of h-BN and TMDs, drag by AFM tips on bulk materials, have load-bearing capacity above $1\mu\text{N}$ and show structural lubricity [52]. Such mechanism might persist even for complex TLs, as those originated by surface wear of these materials via colloidal AFM probes.

## 4. CONCLUSIONS

In summary we addressed the friction response of mesoscopic contacts of $\sim 300\text{nm}$ lateral size by means of colloidal AFM experiments, involving micrometric silica beads sliding on graphite under ambient conditions. We show that after formation of the TL by tribotransferred graphene flakes, contact area reduces to $\sim 50\text{nm}$ (lateral size) and friction response depends on the specific energy landscape experienced by the topographically-highest, triboinduced nanoasperity. At the atomic scale, we find systematic load-controlled transitions from continuous superlubric sliding to stick-slip dissipative motion, that agree with the thermally-activated single-asperity PT model. Vanishing-low dissipative sliding occurs only in specific cases, and below ∼100nN of normally-applied load. Friction force ultimately arises from the interplay between interfacial crystalline incommensurability and pinning effects (both at the contact edges or within the contact area). Random friction fluctuations suggest rearrangements of the tribotransferred flakes triggered by interaction with graphite atomic roughness. Our results point-out the existence of a tight link between mesoscopic graphitic tribopairs and a well-established framework of single-asperity nanotribology. The methodology appears useful to explore mesoscale structural lubricity for other 2D solid lubricants.

## COMPETING INTERESTS

The authors declare no competing financial interests.

## ACKNOWLEDGEMENTS

This work was financially supported by the MIUR PRIN2017 project 20178PZCB5 "UTFROM - Understanding and tuning friction through nanostructure manipulation".

## APPENDIX A. Supplementary data

Supplementary data to this article can be found online at https://doi.org/.

## APPENDIX B. Multiple-tips patterns generated by triboinduced nanocontacts

To simulate the multiple-tip patterns induced by the triboinduced nanoasperities of selected $\mathrm{SiO_2 - TL}$ colloidal probes, we implemented tip convolution of HOPG synthetic surfaces using the morphological dilation algorithm of the open-source software Gwyddion [39]. The tip topography T consisted in a $151 \times 151$ pixels subset extracted from the reverse AFM image of a given TL (Fig. 2c). The synthetic HOPG topography S was a $1000 \times 1000$ pixels map, obtained by superposition of random noise (mimicking 18pm RMS corrugation of the graphite basal plane) with a square based spike [38]. The latter represents a local source of nanoscale or atomic-scale roughness, e.g. associated to graphite multilayer stacks or environmental contamination. Multiple-tip patterns appeared in the dilated topography $\mathrm{I} = \mathrm{S} \oplus \mathrm{T}$ whenever the spike height $h$ was in the range of the TL roughness (Fig. B.1).

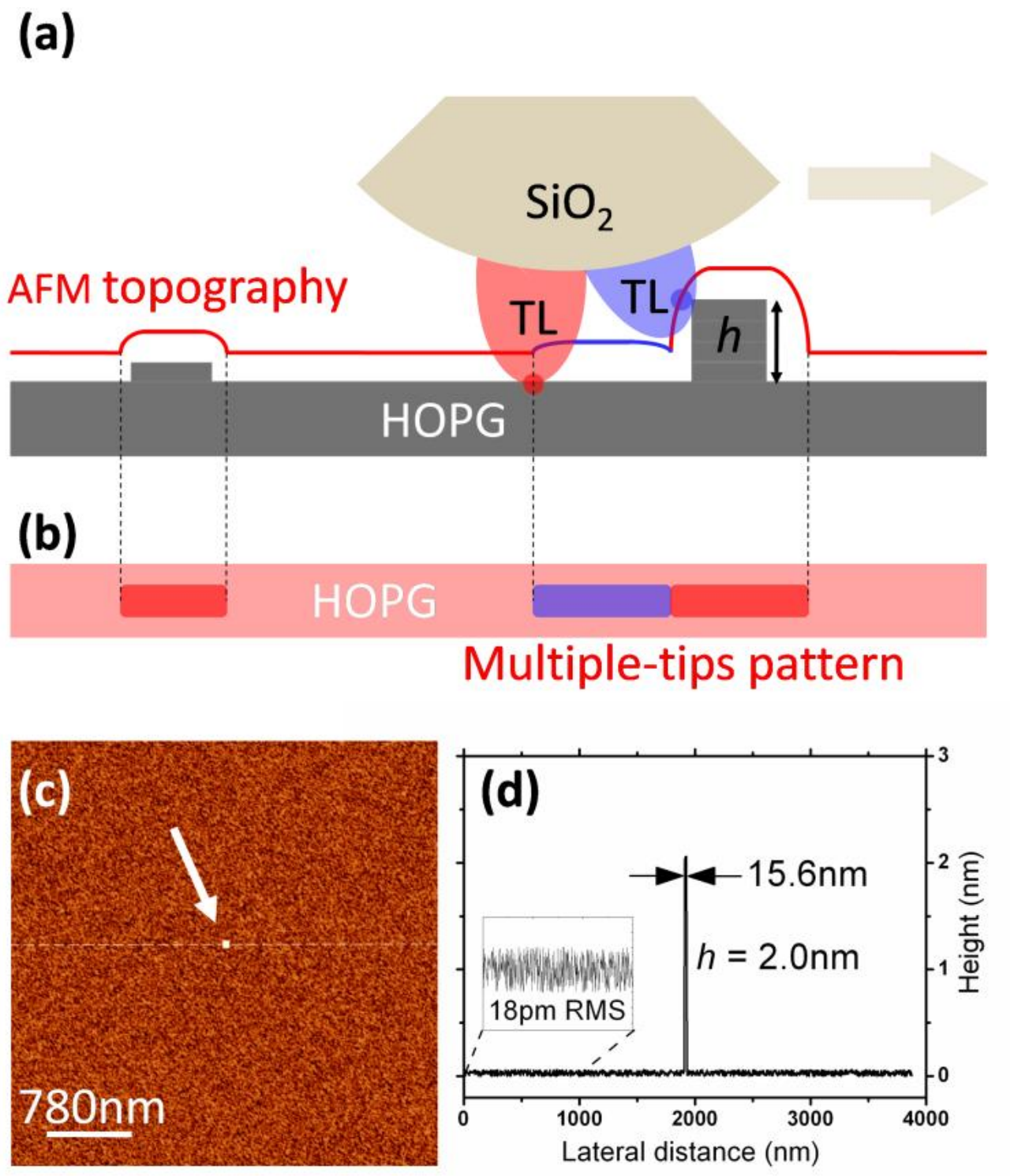

**Fig. B.1.** a) Schematics of a $SiO_2 - TL$ colloidal probe sliding on HOPG: multiple-tips artifacts occur only at the highest surface defect. (b) HOPG topography (top view): appearance of a multiple-tips pattern. (c) Synthetic HOPG map with an isolated, 2nm-tall roughness spike at the centre. (d) Cross sectional height along the dashed line in (c).

Before each dilation process, we intentionally tilted the tip map T to account for small (uncontrolled) variations in the angle formed by the cantilever's long axis with the plane of the sample surface. In fact, such angular variations likely occurred in experiments when replacing the TGT01 tip characterizer grating – providing the colloidal probe topography – with the HOPG substrate on the AFM sample holder. We show below that the tilt of the T map greatly affects the relative positions of the probe nanoasperities, hence the geometry of the patterns in the final dilated topography. Tilt was iterated in small steps ($\pm 0.2°$) until the multiple-tip patterns in the dilated topography $I = S \oplus T$ matched those in real AFM maps (Fig. B.2).

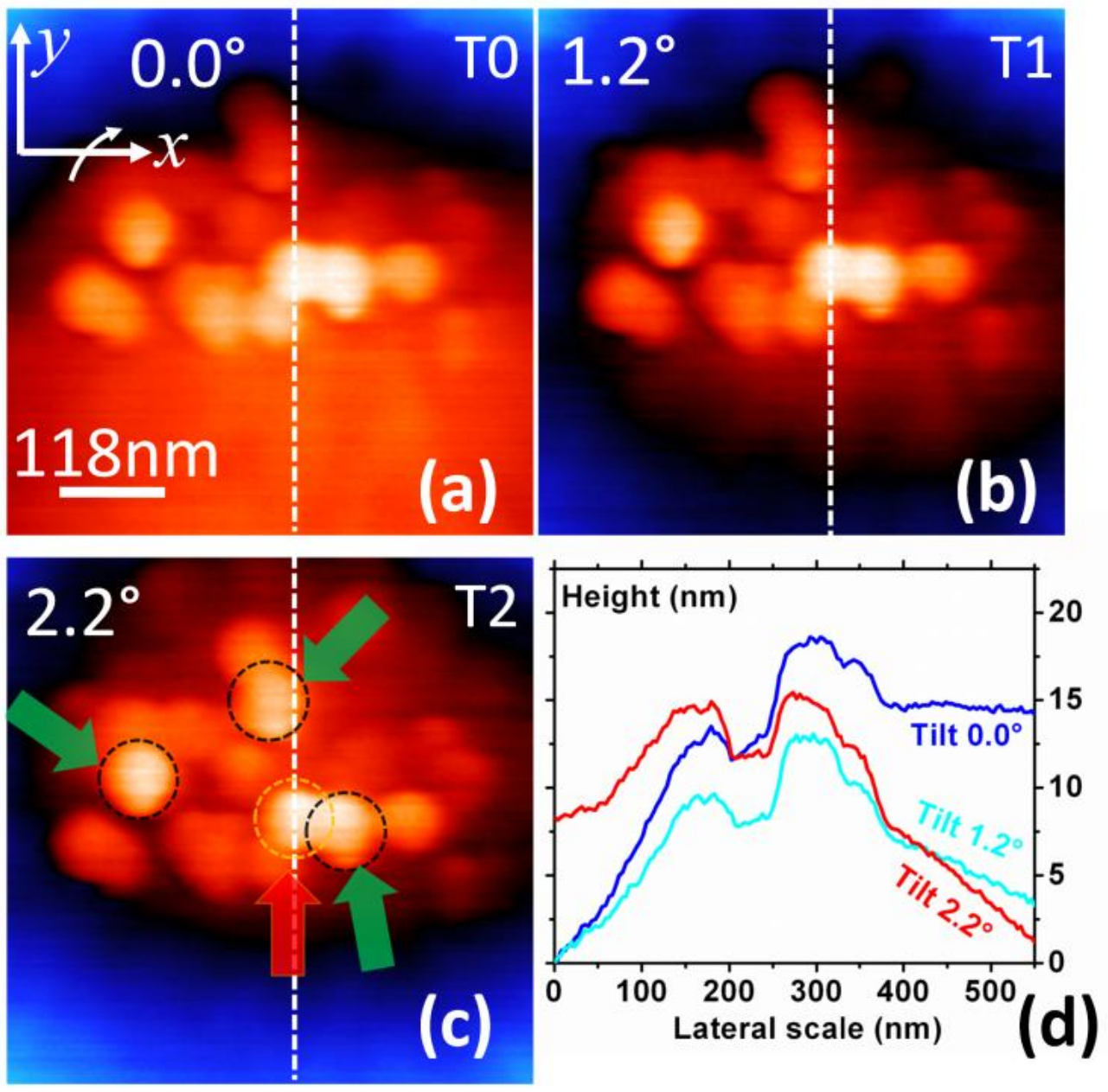

**Fig. B.2.** (a) Reverse AFM topography T0 of the $SiO_2 - TL$ probe in Fig. 2. The cantilever's long axis is aligned with the *y* axis. (b) Topography T1, obtained by applying a tilt of 1.2° to T0 along *y*. (c) Topography T2, obtained by applying a tilt of 2.2° to T0. The main asperities contributing to contact mechanics are circled. The yellow circle highlights the highest asperity making contact on atomically-flat graphite (arrows correspond to those in Fig. 3c). (d) Cross sectional heights along the dashed lines in (a)-(c).

Figure B.3 a-d shows the multiple-tips patterns predicted for a roughness spike of nominal height $h = 2\text{nm}$. Results are reported for different tilts and demonstrate that a 2.2° tilt compensation allows to reproduce the triangular-shaped patterns observed in experiments (Fig. 3a-b). Indeed we found similar patterns for height variations $h = 1 - 5\text{nm}$ (Fig. B.3e), which explains why the triangular shape was robust and ubiquitous over different spots of the HOPG surface (Fig. 3a). Interestingly, four superlubric asperities do interact with the surface spike to generate such pattern. They are highlighted by circles in Fig. B.2c. We underline that multiple-tips patterns were still appreciable for $h = 0.3\text{nm}$ (not shown). However, for $h \leq 0.1\text{nm}$, only an isolated dot appears in the dilated topography. This signals that only the topographically highest asperity of the colloidal probe contacts the sample surface in case of atomic roughness. Such an asperity is highlighted by the yellow circle in Fig. B.2c. This conclusion supports single-asperity modeling of the $SiO_2 - TL/HOPG$ contact when the basal plane of graphite is involved.

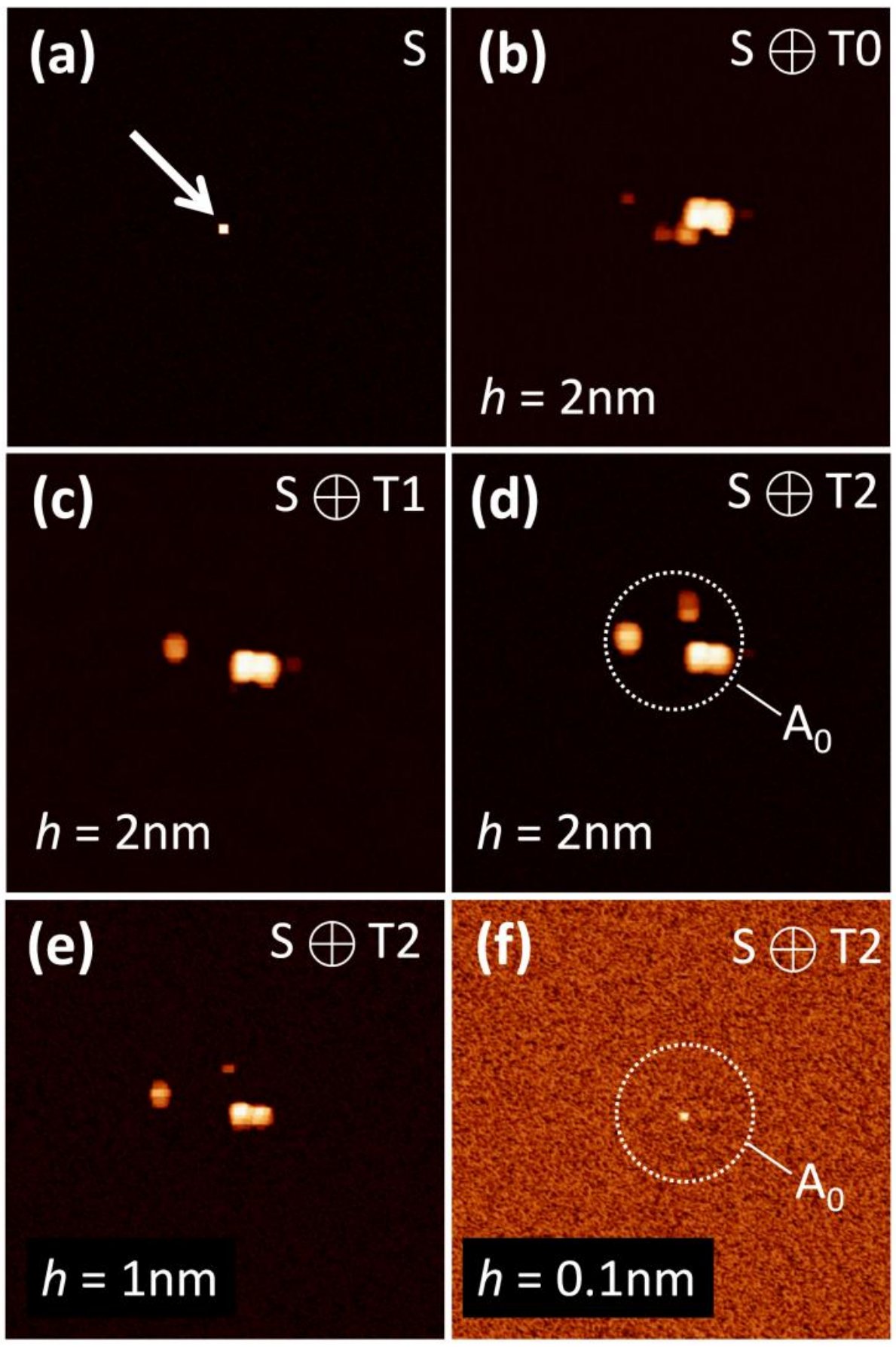

**Fig. B.3.** (a) Synthetic HOPG map S with a roughness spike of height $h$ at the centre (size $950 \times 950\text{nm}^2$). (b) Dilated topography $\text{I} = \text{S} \oplus \text{T0}$ for $h = 2\text{nm}$. The multiple-tips pattern appears at the centre, due to interaction of the surface spike with different protruding asperities on the colloidal probe. (c) Dilated topography $\text{I} = \text{S} \oplus \text{T1}$ for $h = 2\text{nm}$, showing the effect of 1.2° tilt. (d) Dilated topography $\text{I} = \text{S} \oplus \text{T2}$ for $h = 2\text{nm}$, showing that 2.2° tilt gives the triangular shaped pattern observed in experiments. (e) Dilated topography $\text{I} = \text{S} \oplus \text{T2}$ for $h = 1\text{nm}$, still showing the same (but weaker) pattern of (d). (f) Dilated topography $\text{I} = \text{S} \oplus \text{T2}$ for $h = 0.1\text{nm}$. Only an isolated dot appears, since the surface spike now interacts exclusively with one asperity of the colloidal probe (i.e. the highest one). Comparison with $A_0 = 7 \times 10^4\text{nm}^2$ (from Table 1) signals the transformation of the initially mesoscopic contact area into a nanoscopic one.

# Graphite superlubricity enabled by triboinduced nanocontacts

Renato Buzio[1]*, Andrea Gerbi[1], Cristina Bernini[1], Luca Repetto[2], Andrea Vanossi[3,4]

[1]CNR-SPIN, C.so F.M. Perrone 24, 16152 Genova, Italy

[2]Dipartimento di Fisica, Università degli Studi di Genova, Via Dodecaneso 33, 16146 Genova, Italy

[3]International School for Advanced Studies (SISSA), Via Bonomea 265, 34136 Trieste, Italy

[4]CNR-IOM Democritos National Simulation Center, Via Bonomea 265, 34136 Trieste, Italy

*contact author: renato.buzio@spin.cnr.it

# Supplementary data

## S1. Highly Oriented Pirolytic Graphite HOPG

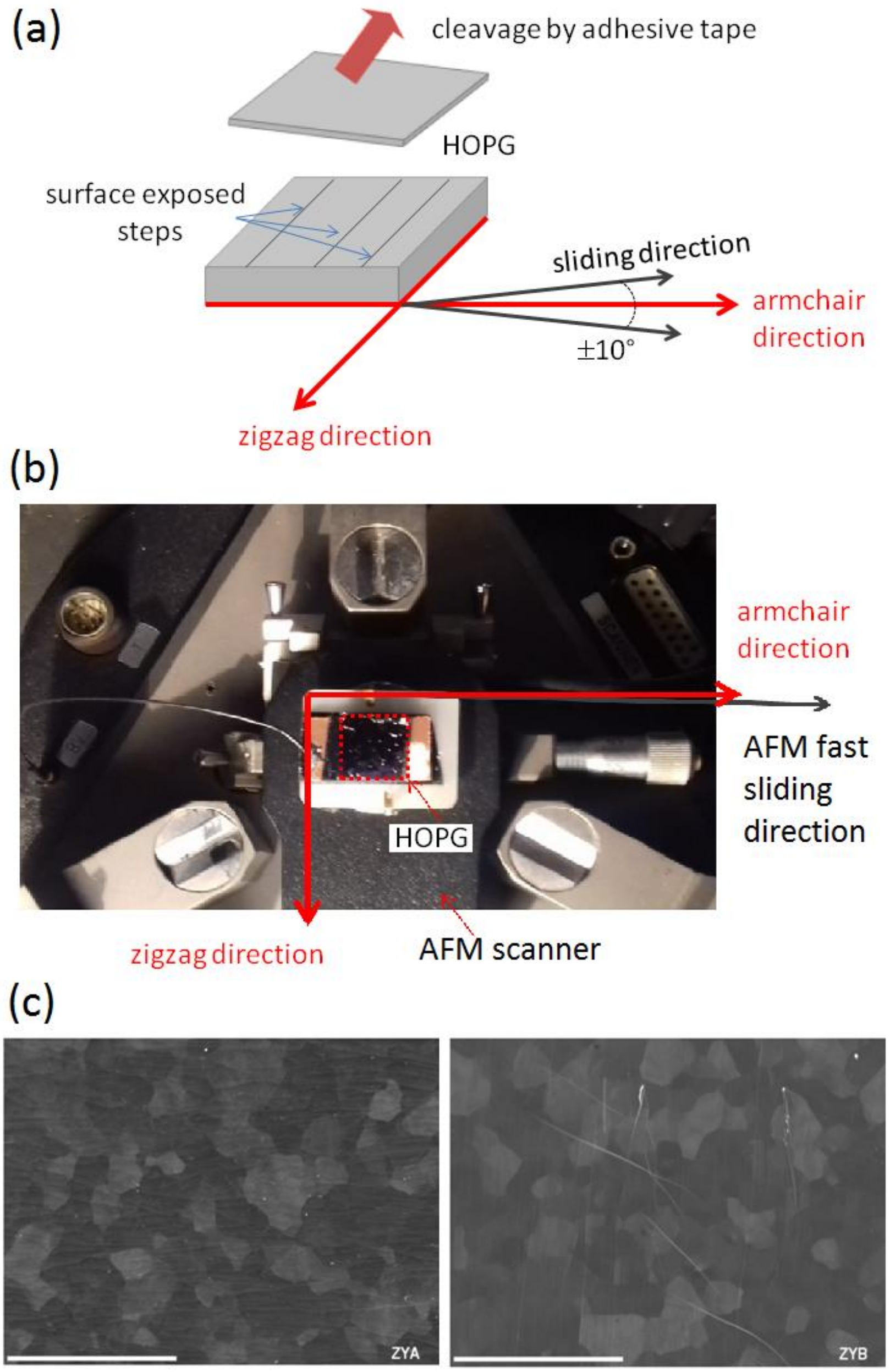

**Figure S.1** (a) Cartoon showing the orientation of the square HOPG substrate with respect to the AFM fast sliding direction. Mechanical cleavage by adhesive tape, carried out along one of the sides of the square HOPG, usually resulted in a preferential orientation of the surface exposed steps along this edge. AFM sliding direction was chosen almost perpendicular (≤ 10°) to this side. Lattice-resolved friction maps allowed *a posteriori* to identify the crystalline directions along the two edges of the square substrate. (b) Top-view photograph of the AFM sample holder. The square HOPG is mounted with parallel sides onto a rectangular-shaped sapphire sample plate (white). The plate is clamped on the scanner by metal clips. This way, AFM sliding direction is kept almost parallel to the armchair direction (misalignment ≤ 10°). (c) Typical SEM images of crystallites in two HOPG samples of different crystalline quality. The scale bar corresponds to 25μm.

## S2. Force calibration and single-asperity modeling

For calibration of normal forces, the sensitivity $s_N$ (in nm/nA) was obtained from the slope of normal deflection *vs* displacement curves acquired over a rigid Si wafer. To obtain the effective spring constant of the colloidal probe $k_N^*$, we first evaluated the normal spring constant $k_C$ of the bare silicon cantilevers (*i.e.* before attaching the silica bead) using Sader's method [1]. Next, we estimated the spring constant $k_N = k_C(L_C/L_B)^3$ , where $L_B$ is the distance of the glued bead with respect to the base of the lever (Fig. S2).

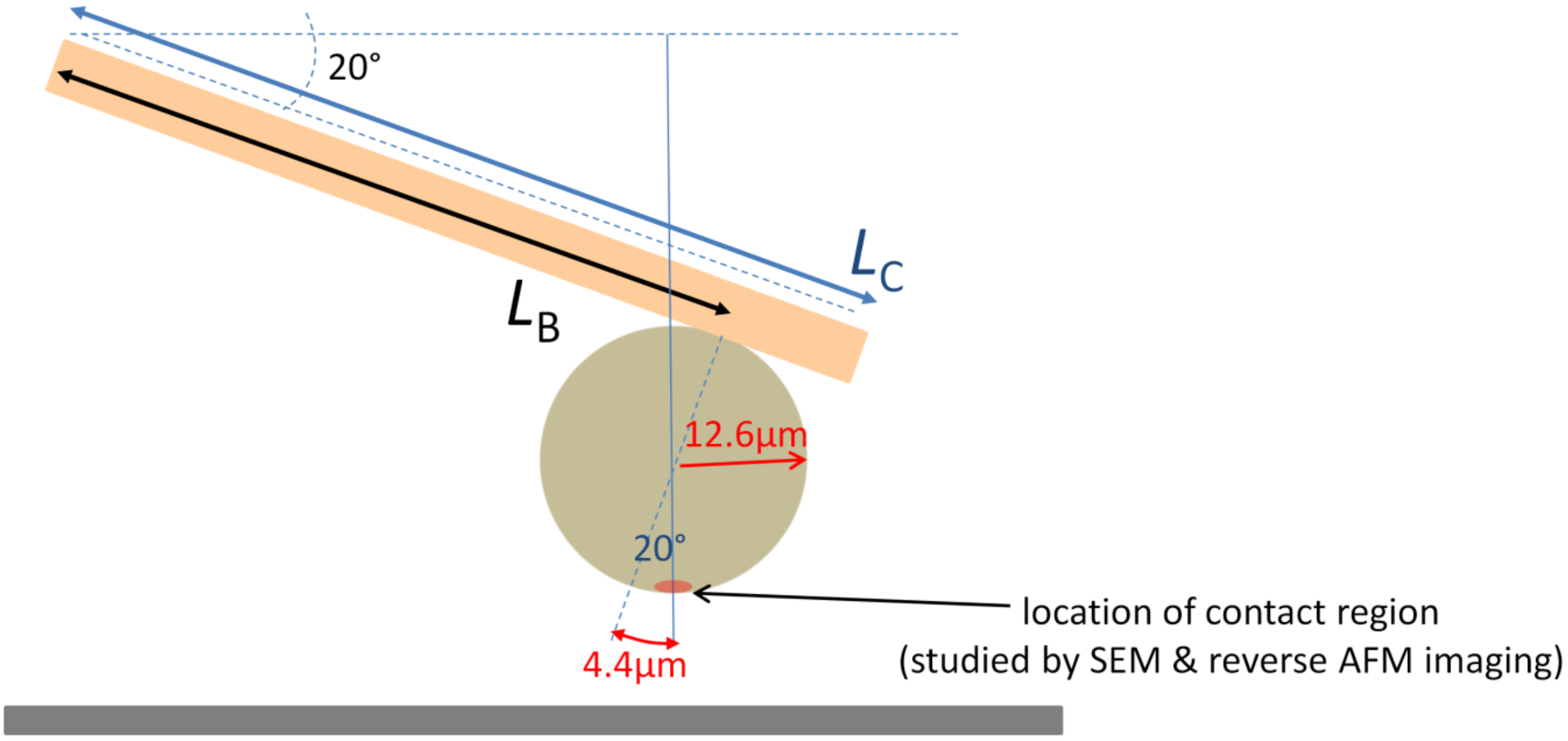

**Figure S.2** Side view of the colloidal AFM probe during AFM experiments. The angle of 20°, formed by the cantilever's long axis with the plane of the sample surface, is imposed by the AFM head design. This angle is taken into account for the calibration of normal forces as well as for the exact location of the contact region in SEM imaging.

Finally, following Edwards *et al.* [2], the effective spring constant $k_N^*$ of the colloidal probe was obtained as:

$$k_N^* \equiv k_N \frac{1}{\cos^2\theta} \frac{1}{\left(1 - \frac{3R}{2L_B} tan\theta\right)} \qquad \text{Eq. (A1)}$$

where $\theta \approx 20°$ is the cantilever tilt with respect to the sample surface, and $R = 12.6\mu m$ is the nominal bead radius. The normal force $F_N$ was thus calculated as $F_N = \alpha_N \Delta U_N = k_N^* s_N \Delta U_N$.

The calibration of lateral force values $F_L$ was accomplished by means of a diamagnetic levitation spring system of lateral stiffness ~82pN/nm (levitated mass $m^L \approx 44.0$ mg; $\omega_d(x_1) \approx 43.095$ rad/s and $\omega_d(x_2) \approx 44.267$ rad/s; $k_{11}(x_1) \approx 81.7$ pN/nm; see [3] for details and notation). Linear interpolation of plots of the lateral deflection signal $\Delta U_L$ (in nA) against the lateral spring displacement $x_1$ (hence the spring force $k_{11}x_1$, in nN) provided the lateral force calibration factor $\alpha_L$ (in nN/nA). The lateral force $F_L$ was thus calculated as $F_L = \alpha_L \Delta U_L$.

In the present study, $\alpha_N \sim 400 - 500\text{nN/nA}$ and $\alpha_L \sim 2\mu\text{N/nA}$, depending on the actual properties of the tipless cantilever and of the bead position along the long axis of the cantilever.

We fitted experimental $F_f$ *vs* $F_N$ curves with the equation A2:

$$F_L = a^2 \left( \frac{\alpha + \sqrt{1 - \frac{F_N}{F_A}}}{1+\alpha} \right)^{4/3} \qquad \text{Eq. (A2)}$$

where $a, \alpha$ are fitting parameters (for notation and formulas see [4]). The mean adhesion force was measured independently from force-distance curves and $F_A$ was constrained to this value for the fit. From the fitted value of the $\alpha$ parameter we evaluated the dimensionless parameter $\lambda$ as $\lambda = -0.924 \ln(1 - 1.02\alpha)$, which is valid for $\alpha < 0.980$. For $\alpha > 0.980$ we simply assumed the JKR limit, $\lambda = \infty$. Given the value of $\lambda$, both the dimensionless critical load $\hat{L}_C$ and the dimensionless contact radius at zero load $\hat{a}_0$ were determined using the following equations:

$$\hat{L}_C = -\frac{7}{4} + \frac{1}{4}\left(\frac{4.04\lambda^{1.4} - 1}{4.04\lambda^{1.4} + 1}\right) \qquad \text{Eq. (A3)}$$

$$\hat{a}_0 = 1.54 + 0.279\left(\frac{2.28\lambda^{1.3} - 1}{2.28\lambda^{1.3} + 1}\right) \qquad \text{Eq. (A4)}$$

In the JKR limit, $\hat{L}_C = -3/2$ and $\hat{a}_0 \sim 1.82$. The interfacial energy $\gamma$ of the contact was estimated as:

$$\gamma = \frac{F_A}{\pi R \hat{L}_C} \qquad \text{Eq. (A5)}$$

where $R$ is the nominal curvature radius of the AFM probe. The contact radius $a_0$ and contact area $A_0$ at zero load were evaluated respectively as:

$$a_0 = \frac{\hat{a}_0}{(K/\pi\gamma R^2)^{1/3}} \qquad \text{Eq. (A6)}$$

$$A_0 = \pi a_0^2 \qquad \text{Eq. (A7)}$$

where $K$ is the combined elastic modulus of tip and sample. Finally the interfacial shear strength was determinate as:

$$\tau = a^2/A_0 \qquad \text{Eq. (A8)}$$

## S3. Transfer layer formation

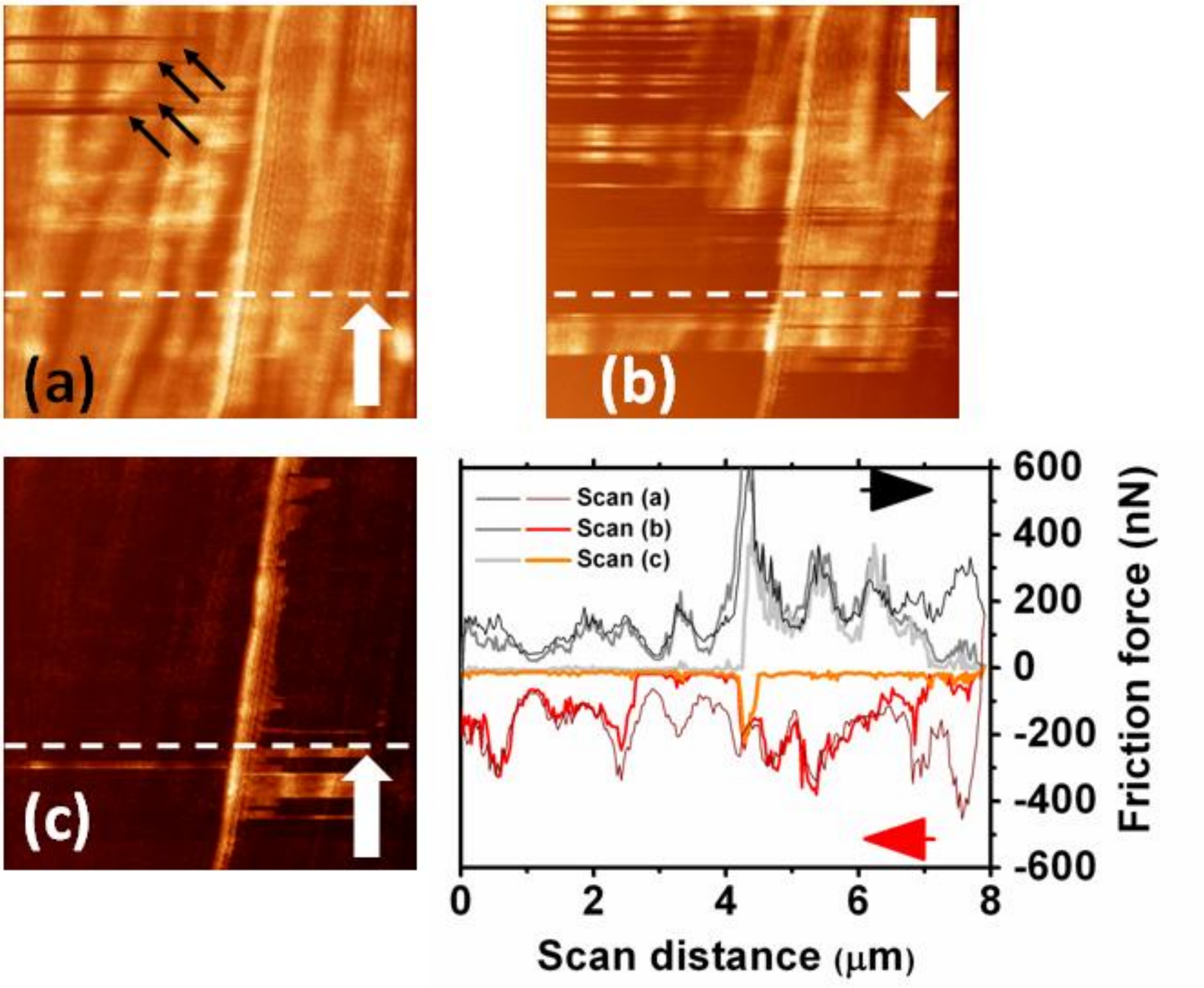

**Figure S.3** (a)-(c). Friction maps acquired sequentially over the same HOPG region, immediately after the acquisition of the friction map in Fig. 1a (see main text). The white arrow is the slow scan-direction for each map. Black arrows in (a) highlight sudden transitions to the 'superlow-friction' state. (d) Friction loops corresponding to the dashed lines in (a)-(c).

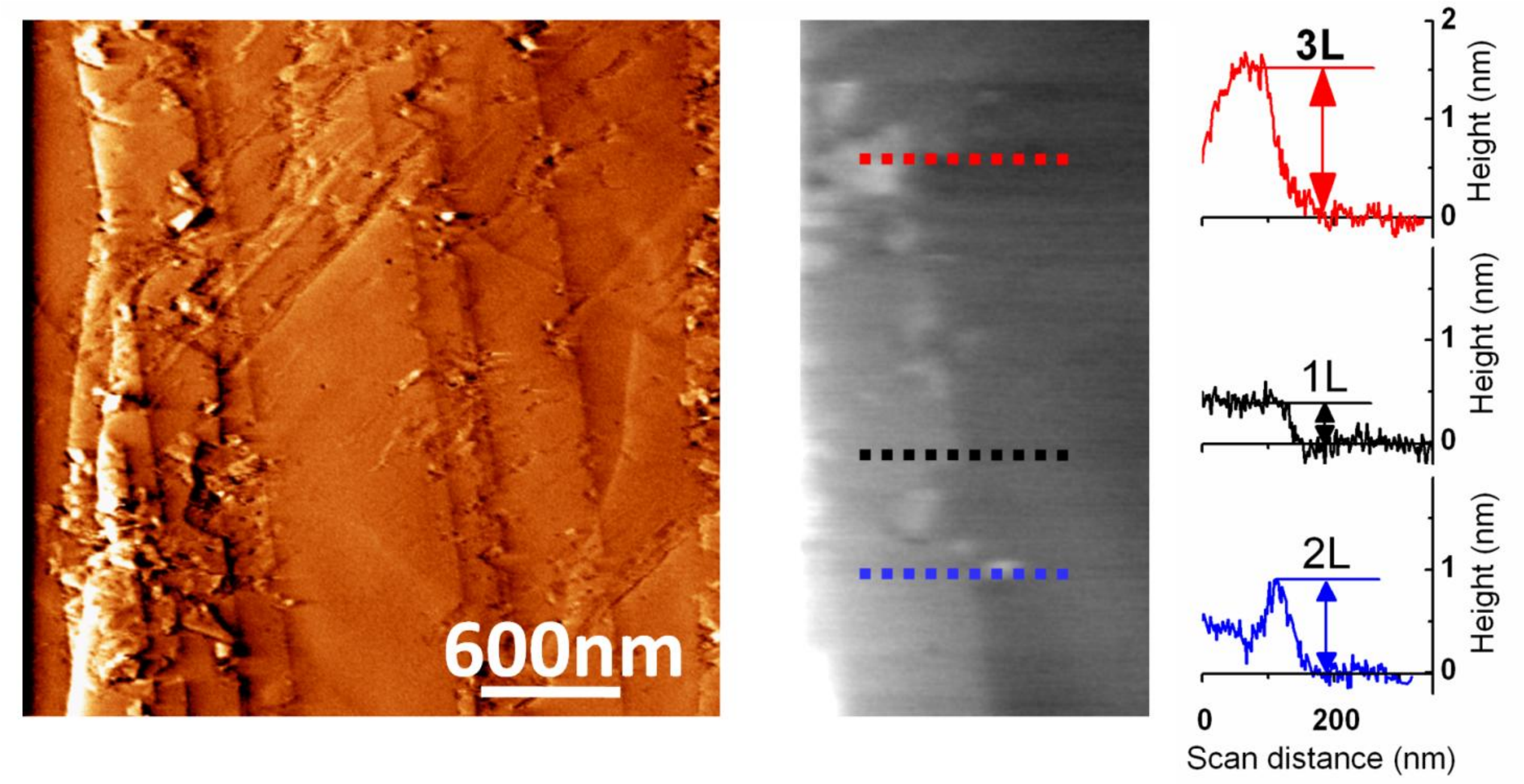

**Figure S.4** (Left) *Ex-post* AFM map showing HOPG abrasive wear, specifically worn graphite monolayers and debris, after colloidal probe sliding across step edges and TL formation. (Right) Magnified topography and selected cross-sections highlight graphite exfoliation at a step edge with generation of nanoflakes ($1L \approx 0.5nm; 2L \approx 1.0nm; 3L \approx 1.5nm$). Few-layers graphene flakes, with lateral size of tens of nm, find excellent correspondence with the tribotransferred flakes directly imaged by reverse AFM and SEM on top of the silica beads.

## S4. Adhesion at the colloidal probe - HOPG interface in ambient conditions

The adhesion force at the silica-HOPG interface can be calculated using analytical approximations for the capillary and vdW forces, under the condition of undeformable surfaces. To this end the surface roughness of the pristine silica beads is explicitly taken into consideration. A representative AFM morphology of a silica bead is reported in Fig. S.5a,b. Figure S.5c shows that the related surface height distribution is nearly Gaussian, with standard deviation $\sigma \sim 0.5$nm. HOPG in turn is treated as an ideally-smooth surface, its roughness being comparable to or smaller than the noise level of the AFM ($\sigma \sim 15$pm).

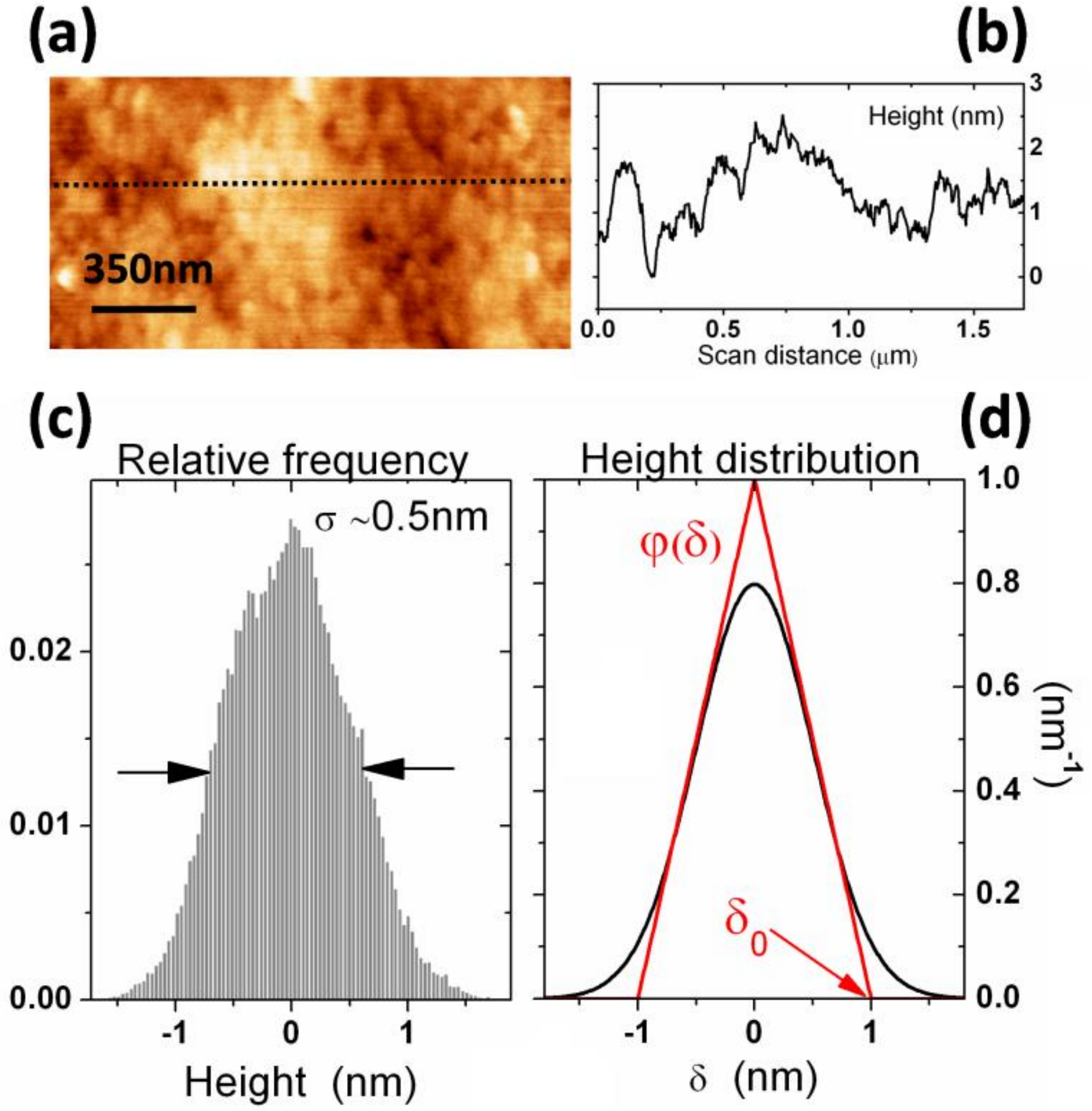

**Figure S.5** (a) AFM topography (after parabolic flattening) of the contact region of a pristine silica microsphere. (b) Height profile along the dashed line in (a). (c) Histogram of heights extracted from the topography in (a). (d) Comparison between the Gaussian probability distribution for surface heights (standard deviation $\sigma = 0.5$nm; black) and the triangular approximation used to calculate the capillary force ($\delta_0 = 1$nm; red).

Capillary forces are known to dominate over the other surface forces whenever the contact involves hydrophilic surfaces and occurs under ambient conditions. First of all we focus on the magnitude of the capillary adhesion between the pristine silica microsphere and HOPG. According to Farshchi-Tabrizi *et al.* [5], the capillary force $F_{cap}$ of a water meniscus condensed around a sphere-on-flat contact geometry with ideally-smooth surfaces can be calculated as $F_{cap}(D) = 2\pi\gamma_S R(2c - D/r)$, where $D$ is the distance between the sphere of radius $R$ and the flat, $\gamma_S = 0.072$ J/m$^2$ is the water surface tension, $r = -\lambda_k/\ln(P/P_0)$ is the (meridional) curvature radius of the liquid meniscus related to the water Kelvin length $\lambda_k = 0.523$nm and the ambient relative humidity $P/P_0$, $c = (cos\theta_1 + cos\theta_2)/2$ is the average cosine of contact angles of water with respect to the two involved surfaces (silica and HOPG). The contact angle $\theta_1$ for the (hydrophilic) silica microspheres can be safely assumed to be in the range between 0° and 35° [6]; for HOPG we choose $\theta_2 \sim 60°$ [7]. This results in $c \sim 0.66 - 0.75$ and the capillary force $F_{cap}(D = 0) = 4\pi\gamma_S Rc \sim 7.5 - 8.6$μN, that greatly exceeds the overall adhesion $F_A$ measured experimentally (Fig. 1c). As $F_{cap}(D = 0)$ does not depend on $r$, it turns out that such estimate of the capillary force does not depend on ambient humidity. However, it is well known that even a tiny amount of interfacial roughness in the nm range critically affects the capillary force. Roughness prevents the microsphere to get in intimate contact with the plane over an extended area. Hence contact occurs at the higher nanoasperities, and different stages of capillary condensation take place depending on RH value and the local gap between the two surfaces. We evaluate the capillary force at the silica bead-HOPG interface via the analytical approximation developed by Farshchi-Tabrizi *et al.* [5] for the case of two rough microspheres having a triangular joint distribution of nanoasperity heights $\phi(\delta)$:

$$\phi(\delta) = \begin{cases} \frac{\delta_0+\delta}{\delta_0^2} & -\delta_0 < \delta < 0 \\ \frac{\delta_0-\delta}{\delta_0^2} & 0 < \delta < \delta_0 \\ 0 & \delta < -\delta_0 \text{ and } \delta > \delta_0 \end{cases} \qquad \text{Eq. (A9)}$$

In the present case, HOPG is ideally-smooth compared to silica, therefore the combined distribution of nanoasperity heights matches the Gaussian distribution of surface heights of the silica beads. This in turn can be approximated by the triangular distribution of Eq.(A9), with $\delta_0 = 2\sigma \approx 1$nm (Fig. S.5d). It follows that $F_{cap}$ can be calculated from simple polynomial expressions of the length coordinate $z \equiv 2rc - D$, each expression working for a specific $z$ range:

$$F_{cap}(z) = \begin{cases} \frac{\pi R^*\gamma_S}{3}\frac{z^3}{\delta_0^2 r} & 0 \le z \le \delta_0 \\ \frac{\pi R^*\gamma_S}{3}\frac{[z^3-2(z-\delta_0)^2]}{\delta_0^2 r} & \delta_0 \le z \le 2\delta_0 \\ \pi R^*\gamma_S\frac{2(z-\delta_0)}{r} & 2\delta_0 \le z \le R \end{cases} \qquad \text{Eq. (A10)}$$

The parameter $R^*$ is the effective radius in the two-spheres contact model, and corresponds to the bead radius ($R^* = R = 12.6\mu m$) for the sphere-on-flat contact geometry. Figure S.6a shows how the capillary force predicted from Eq. (A10) varies with the ambient humidity. For the laboratory condition RH~60% and distance $D = 0$, one gets $F_{cap} \sim (2.6 \pm 0.4)\mu N$. The error $\pm 0.4\mu N$ reflects uncertainty in the wetting angle $0° - 35°$ of the silica microbeads.

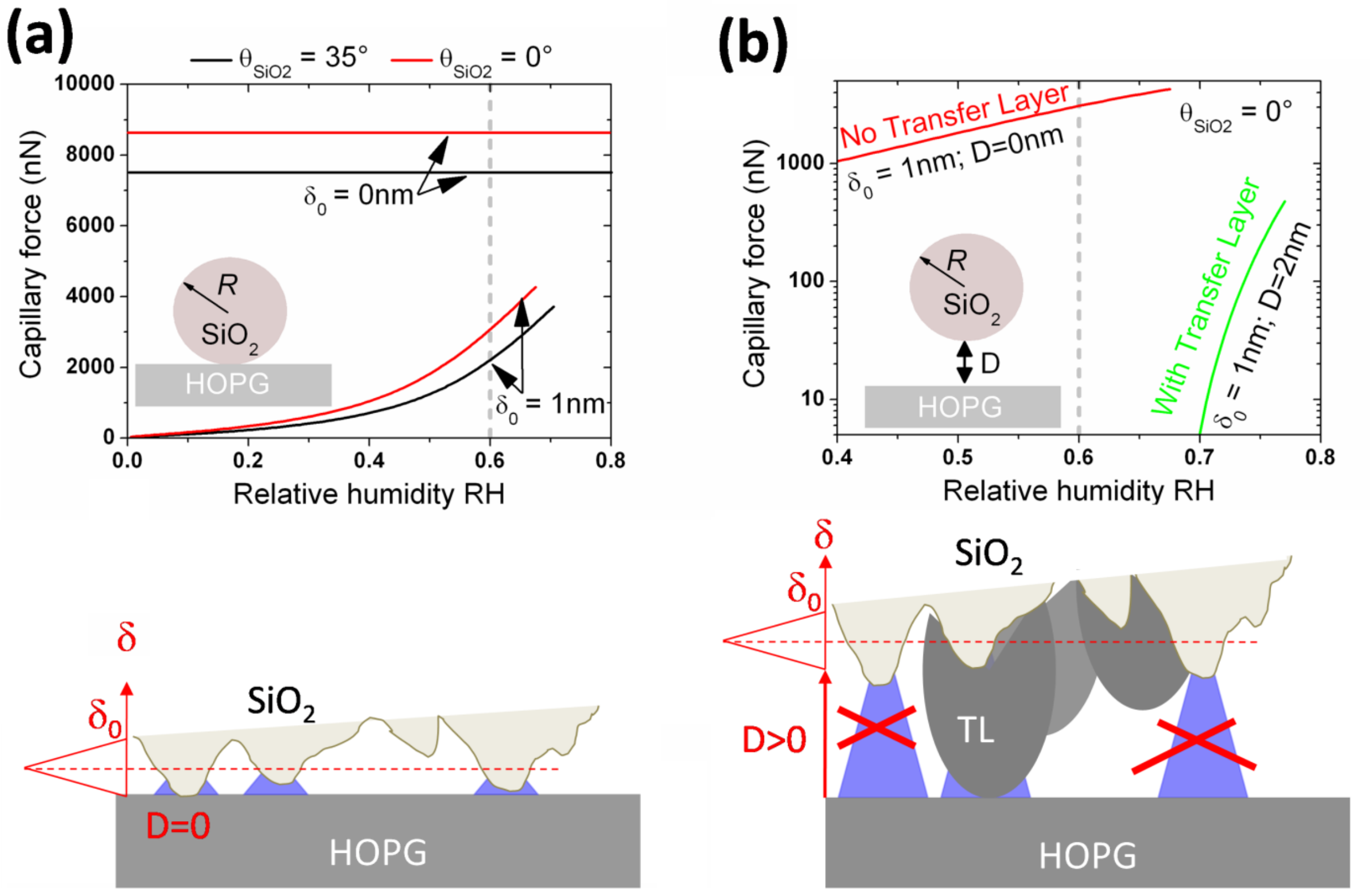

**Figure S.6** (a) Capillary forces predicted by analytical approximations for smooth ($\delta_0 = 0$) and rough ($\delta_0 = 1$nm) silica microbeads in contact with HOPG. Two wetting angles are considered for silica. (b) Capillary forces predicted for a rough microsphere in contact with HOPG, respectively without any interfacial transfer layer ($D = 0$) and with a 2nm-tick one ($D = 2$nm). Cartoons depict a magnified view of the contact region between the bead and HOPG.

Within such framework, the effective cross-sectional contact area between the silica probe and HOPG amounts to:

$$A = \frac{\pi R}{3} \frac{[z^3 - 2(z - \delta_0)^2]}{\delta_0^2} \sim 3.1 - 4.4 \times 10^4 \text{nm}^2 \qquad \text{Eq. (A11)}$$

The area $A$ quantifies those interfacial regions contributing to capillary condensation and formation of liquid menisci. It is only a factor 2 smaller than the contact area $A_0 \sim 7 \times 10^4 \text{nm}^2$ predicted by the JKR model (Table I). Hence, different arguments and contact models consistently confirm that the pristine bead-HOPG contact junction is mesoscopic in size.

In addition to the capillary force, also the long-range vdW force contributes to total adhesion between the pristine microspheres and HOPG. Again, silica interfacial roughness limits the distance of closest approach between the two contacting surfaces to a separation of about $\delta_0$. We estimate vdW attraction for the sphere-on-flat contact geometry as:

$$F_{vdW}(\delta_0) \approx \frac{A_H R}{6(\delta_0+D_0)^2} \sim 209\text{nN} \qquad \text{Eq. (A12)}$$

where $A_H = 1.35 \times 10^{-19}$J is the Hamaker constant for the silica-HOPG interface [8] and $D_0 = 0.17$nm is the interatomic spacing at the contact spot [5]. The (total) estimated adhesion is clearly dominated by capillary forces and amounts to $F_{cap} + F_{vdW} \sim (2.8 \pm 0.4)\mu$N, in very good agreement with $F_A \sim 2 - 3\mu$N measured experimentally.

We now consider the effect of the triboinduced TL on the adhesion force at the silica bead-HOPG interface. As discussed above and shown in Fig. 2, the TL consists of a number of superlubric asperities protruding a few nanometers above the profile of the microbead. Such asperities limit the distance of closest approach between the bead and HOPG. It is convenient to express the total adhesion force as:

$$F_A = F_{Cap}^{micro} + F_{vdW}^{micro} + F_{Cap}^{nano} + F_{vdW}^{nano} \qquad \text{Eq. (A13)}$$

where ‘micro’ contributions refer to the micrometric bead, and ‘nano’ contributions to the TL adhesion effect (see the ‘two-spheres’ model in [5] and Fig. S.7a). For simplicity we focus on a minimal TL model consisting of one main triboinduced tip, of 2nm height. In such case Eq. (A10) predicts that water condenses into the contact gap only when the condition $z = 2cr - D > 0$ is satisfied, with $D = 2$nm. For a silica contact angle $\theta_1 = 0°$ this implies a relative humidity RH higher than ~68%. More precisely, Figure S.6b shows that the 2nm-thick TL fully suppresses the capillary force between the micrometric bead and HOPG at low humidity (RH < 68%), or reduces it by orders of magnitude (1-100nN) at higher humidity (RH ≥ 68%). Hence, $F_{Cap}^{micro} \sim 0$nN at RH = 60%. We claim that this is the main mechanism responsible for the abrupt drop of the adhesion force $F_A$ between the AFM colloidal probes and HOPG when the TL is formed. A 2nm-thick TL greatly reduces the magnitude of the long-ranged vdW forces too, according to the following estimation:

$$F_{vdW}^{micro} = F_{vdW}(\delta_0 + D) \approx \frac{A_H R}{6(\delta_0+D+D_0)^2} = 28\text{nN} \qquad \text{Eq. (A14)}$$

Therefore, whenever a triboinduced TL is formed at the silica-HOPG interface, the only relevant contributions to total adhesion $F_A$ likely come from the capillary force $F_{Cap}^{nano}$ and the vdW force $F_{vdW}^{nano}$ acting between HOPG and the main protruding contact asperity of the TL (Fig. S.7). Estimations of such forces rely on knowledge of fine details of the morphology of the single-asperity contact, as extensively discussed in [5]. Assuming a spherical asperity of curvature radius $R_{nano} = 300$nm, and contact angle $\theta_1 = \theta_2 \sim 60°$ for such graphitic tip, the meniscus force can be calculated as:

$$F_{Cap}^{nano} = 2\pi\gamma_S R_{nano} sin\beta[\sin(\beta+\theta) + \frac{R_{nano}}{2r} sin\beta] \qquad \text{Eq. (A15)}$$

where $\theta = \theta_1 = \theta_2 = 60°$, $\beta$ is the angle describing the position of the contact line of the liquid on the sphere $(0 \leq \beta \leq \cos^{-1}(1 - D/R_{nano}))$ and $r = R_{nano}(1 - cos\beta)/(\cos(\beta+\theta) + cos\theta)$ is the meridional curvature radius of the meniscus (see [5] for details). The vdW adhesion force is instead calculated as:

$$F_{vdW}^{nano} = \frac{A_H^{GR} - A_H^{GR,l}}{6R_{nano}} \frac{D_0/R_{nano} + 1 - 2cos\beta}{(D_0/R_{nano} + 1 - cos\beta)^2} + \frac{A_H^{GR}}{6R_{nano}} \frac{D_0/R_{nano} - 1}{(D_0/R_{nano})^2} - \frac{A_H^{GR}}{6R_{nano}} \frac{D_0/R_{nano} + 2D/R_{nano} - 1}{(D_0/R_{nano} + D/R_{nano})^2} \quad \text{Eq. (A16)}$$

where we assumed a factor ~6 reduction of the vacuum graphite-graphite Hamaker constant $A_H^{GR} = 2.8 \times 10^{-19} J$ [8] due to capillary water, i.e. $A_H^{GR,l} \sim A_H^{GR}/6 \sim 0.5 \times 10^{-19} J$.

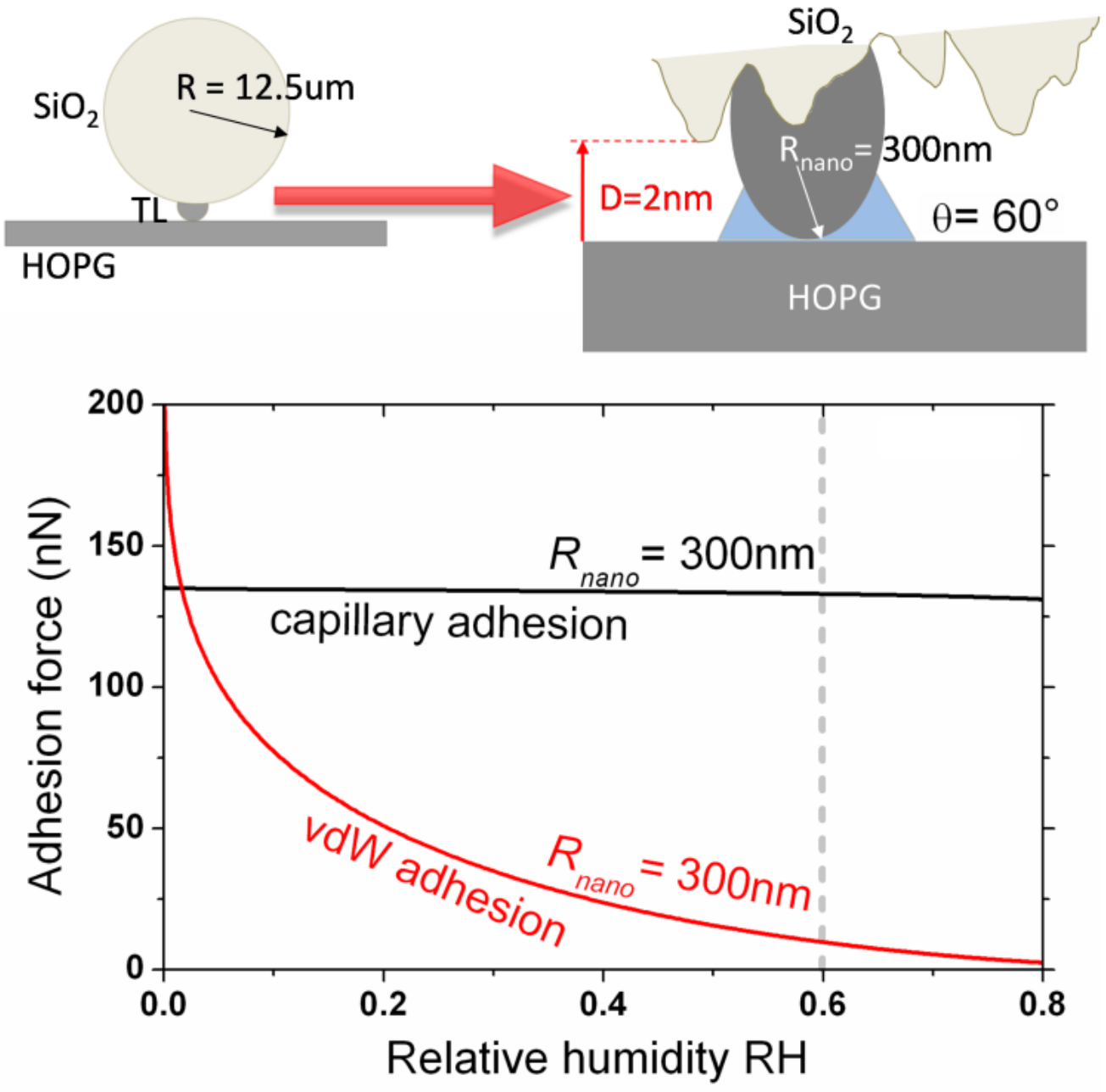

**Figure S.7** (Top) Two-spheres model describing the $SiO_2$/HOPG junction after formation of the triboinduced TL. The latter is treated as a 2nm-tall, graphitic asperity of 300nm curvature radius. The water contact angle is the same for the asperity and HOPG, $\theta = 60°$. (Bottom) Capillary and vdW forces calculated for the spherical contact asperity, using Eqs. (A15),(A16). Contributions from the micrometric sphere are negligible (see text).

Figure S.7 shows how the capillary and vdW forces predicted from Eqs. (A15),(A16) vary with the ambient humidity. For the laboratory condition RH~60% one gets $F_{Cap}^{nano} = 133\text{nN}$ and $F_{vdW}^{nano} \approx 10\text{nN}$.

From previous estimations, the total adhesion force at the $\text{SiO}_2 - \text{TL/HOPG}$ interface is $F_A \approx F_{vdW}^{micro} + F_{Cap}^{nano} + F_{vdW}^{nano} \approx 170\text{nN}$. This agrees very well with $F_A$ measured experimentally ($F_A = 100 - 200\text{nN}$) for the triboinduced 'ultralow-friction' interfaces.

The above description of interfacial adhesion is highly comprehensive, as accounts for variations of interfacial roughness and contact angle due to the formation of the TL. With this approach we gain a sound physical basis for the breakdown of the adhesion $F_A$ mediated by the TL. Nonetheless, interfacial deformations are fully neglected and the actual TL contact geometry is more complex than assumed in calculations. Qualitatively, however, we do not expect that such improvements might affect the main conclusion that $F_{Cap}^{micro} \rightarrow 0$ when interfacial roughness increases by a few nanometers.

**S5. Multiple-tip effects originated by a rough transfer layer**

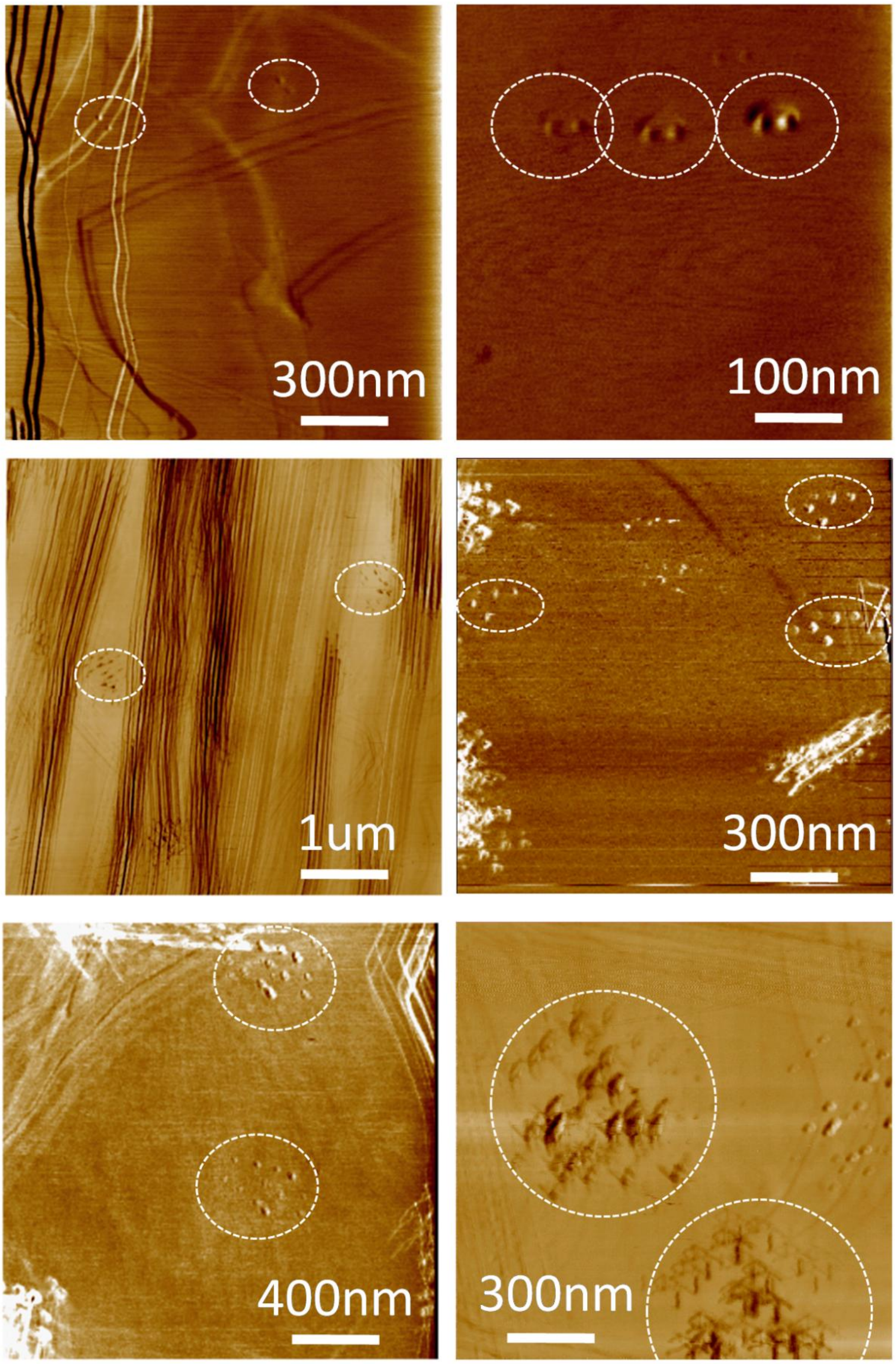

**Figure S.8** Examples of multiple-tip patterns in lateral force maps acquired by $SiO_2$ − TL colloidal probes. Besides such patterns, multiple step edges also occur.

## S6. Atomic-scale friction experiments

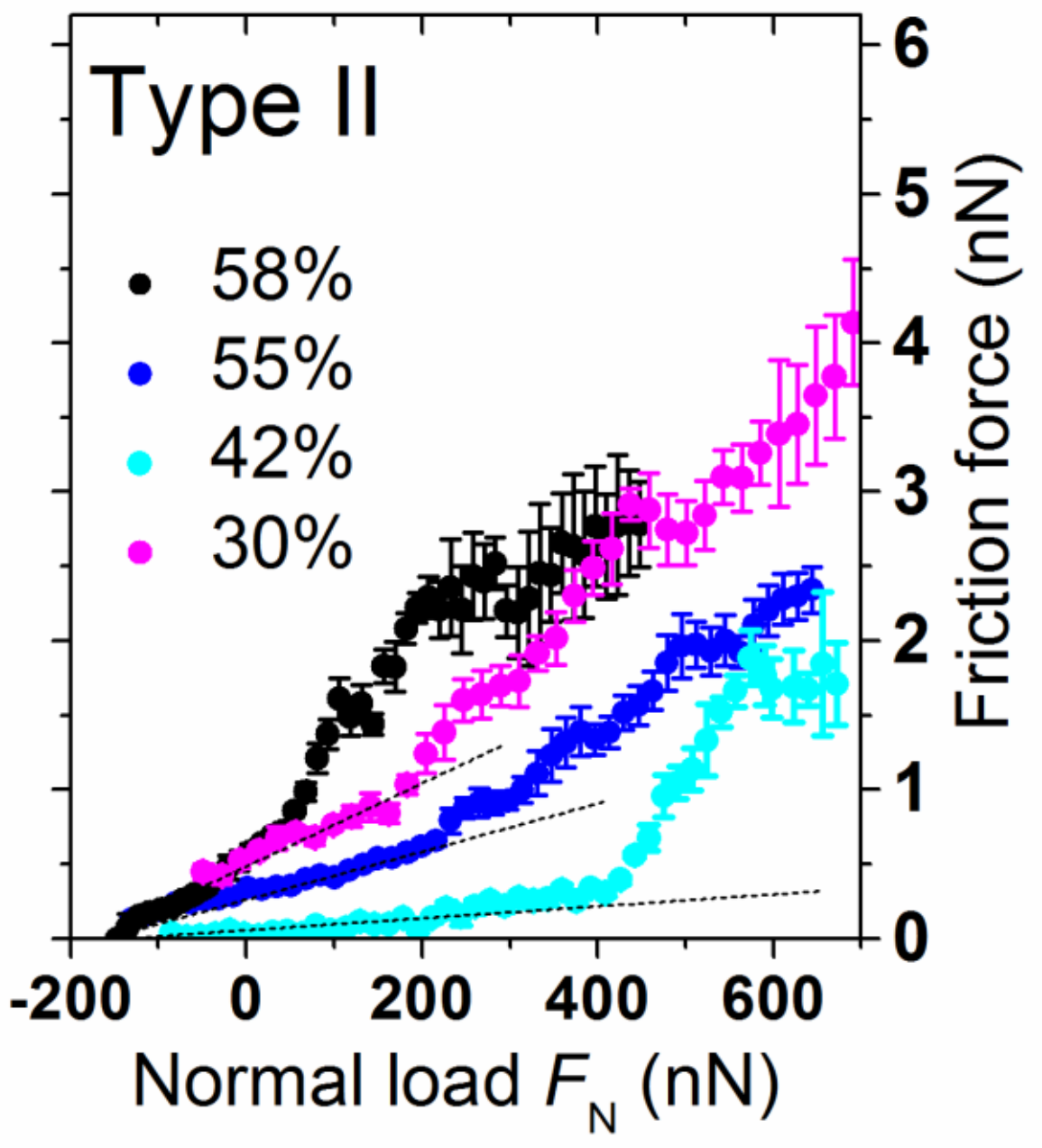

**Figure S.9** Friction *vs* load characteristics for the contact junctions corresponding to the black and pink datasets in Fig.4f (scan area 11x11nm$^2$, scan velocity 33nm/s). The type II characteristics of Fig. 4b are also shown.

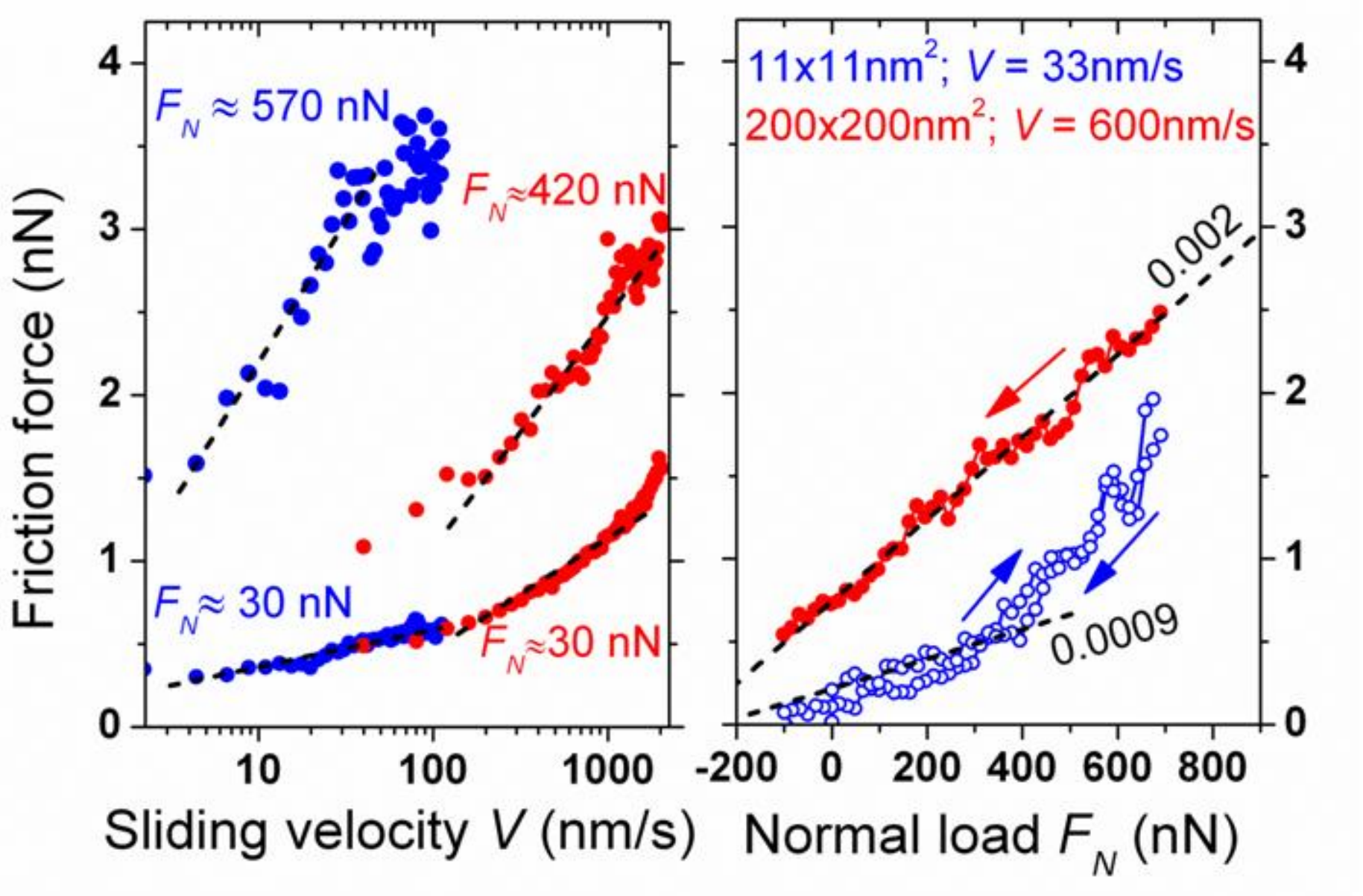

**Figure S.10** Representative friction vs velocity and friction vs load characteristics (left and right panels respectively) for the $SiO_2$-TL/HOPG interface. The same colloidal probe was used in both cases. Blue symbols refer to a data set measured on a 11x11nm$^2$ scan area. Red symbols represent a data set measured on a 200x200nm$^2$ area (including the smaller one). The contact junction clearly evolved between the two data sets, resulting in two different dependencies of friction force on normal load (type I *vs* type II).

## S7. Numerical solutions to the thermally-activated, one-dimensional PT model

In the main text, experimental $F_f$ $vs$ $F_N$ characteristics are transformed into $F_f^*$ $vs$ $\eta$ curves, where $F_f^* \equiv F_f/ak$ is the adimensional friction force and $\eta$ is the Tomlinson parameter. As discussed below, this choice makes easier to compare the 1D PT model with experimental contact junctions having different, load-dependent contact parameters $E_0$ and $k$.

In principle, any numerical solution of the Langevin equation requires to chose specific values for the contact stiffness $k$, the corrugation amplitude $E_0$ and the mass $m$. From the theoretical point of view, at T=0K and in the quasi-static limit $v \to 0$, the $F_f^*$ $vs$ $\eta$ relationship can be approximated by the analytical expression $F_f^* = \frac{1}{2\pi}\left(\eta - \pi + \left(\frac{4}{3}\right)\sqrt{\pi/\eta} - 1/2\eta\right)$ which is known to work well for $\eta > 2$ [9]. According to Fig. S10, numerical solutions of the Langevin equation calculated for $\mathrm{T} = 0\mathrm{K}$ and $v = 30\mathrm{nm/s}$ perfectly agree with such expression. Figure S11 also shows numerical solutions for T=296K and $v = 30\mathrm{nm/s}$. Thermolubricity makes the $F_f^*$ $vs$ $\eta$ curves to lie systematically below their athermal limit. Additionally, solutions calculated for different $k$ values (15N/m, 20N/m, 25N/m) split into three separate curves. The vertical dispersion is however small (< 5%) over the whole range of $\eta < 10$. Hence, for the purpose of comparison with experiments, the $F_f^*$ $vs$ $\eta$ characteristics at $\mathrm{T} = 296\mathrm{K}$ can be assumed to be $k$-independent to a good approximation.

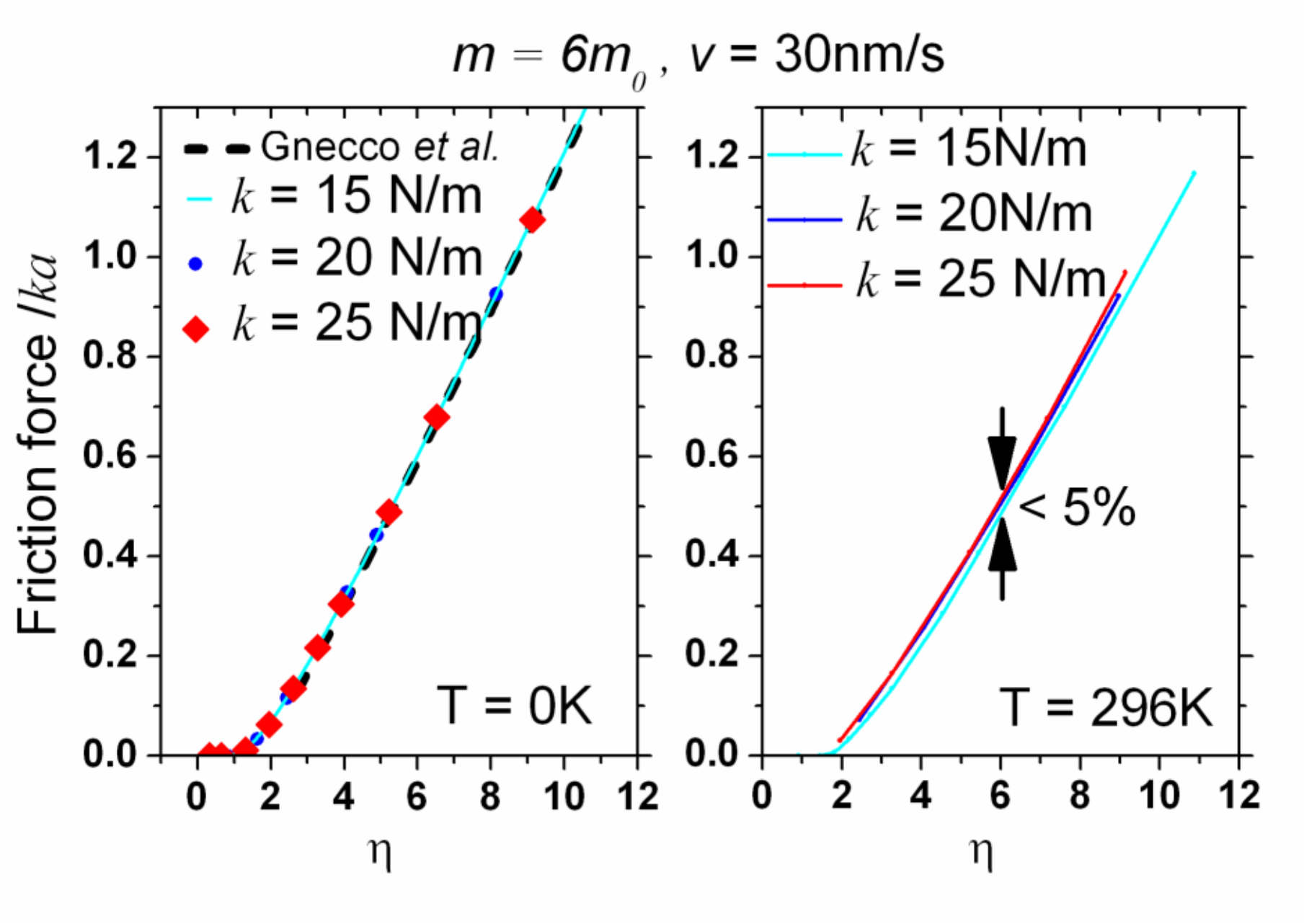

**Figure S.11** Numerical solutions to Langevin equation for atomic friction, calculated with mass $m = 6m_0$ ($m_0 = 1 \times 10^{-12}$Kg) and Langevin damping $\gamma = 0.01\ ns^{-1}$. The analytical expression developed by Gnecco et al. [9] is also reported.

Accordingly, the mass parameter $m$ was adjust in order to obtain the best agreement between experimental data and numerical solutions. The generated $F_f^*$ $vs$ $\eta$ curves were compared with experimental data by visual inspection, and $m$ was adjusted to minimize the difference between experimental and predicted values. A satisfying result was found for $m$ in the range between $m_0$ and $6m_0$. Such $m$ values agrees with those typically reported in literature ($\sim 10^{-12} - 10^{-10}$Kg , e.g. see [10,11]).

## Supplementary references